\documentclass[a4paper,11pt]{article}
\usepackage{jcappub} 
\usepackage{lineno}
\usepackage[T1]{fontenc}
\usepackage{gensymb}
\usepackage{subcaption}
\usepackage{multirow}
\usepackage{graphicx}

\title{\boldmath Cosmological redshift of a Schwarzschild-de Sitter black hole: \\ 
Towards estimating the Hubble constant}

\author[a,1]{D. Villaraos,\note{Corresponding author.}}
\author[a]{A. Herrera-Aguilar,}
\author[b]{M. Momennia}
\author[b]{and U. Nucamendi}

\affiliation[a]{Instituto de F\'\i{}sica, Benem\'{e}rita Universidad Aut\'{o}noma de Puebla\\Av. San Claudio, Ciudad Universitaria, 72570, Puebla, Puebla, M\'exico}
\affiliation[b]{Facultad de Ciencias F\'\i{}sico-Matem\'{a}ticas, Universidad Michoacana de San Nicol\'{a}s de Hidalgo,\\Ciudad Universitaria, 58040, Morelia, Michoac\'{a}n, M\'{e}xico}

\emailAdd{deborahv@ifuap.buap.mx}

\abstract{In this work we estimate the parameters of several astrophysical black holes hosted at the core of active galactic nuclei by studying the kinematics of test objects in their accretion disk. First, we derive expressions for the redshift and blueshift of photons emitted by a massive particle circularly orbiting a Schwarzschild-de Sitter black hole, and detected by a distant receding observer. The frequency-shift depends on the mass and distance of the black hole, the orbital radius of the photon source, as well as the Hubble constant, directly relating these quantities to astrophysical observables, namely, the redshift and the angular position of the emitting particle on the sky. We apply for the first time this theoretical model, which accounts for the universe expansion through the Schwarzschild-de Sitter metric, to real astrophysical systems using megamaser galaxies within the Hubble flow, namely UGC 3789, NGC 5765b, NGC 6264, NGC 6323, and CGCG 074-064. Bayesian inference based on Markov Chain Monte Carlo methods is employed to estimate the mass-to-distance ratio, the product of the Hubble constant with the distance, and the black hole angular position. Additionally, by assuming a Gaussian prior on the Hubble constant, the mass, distance, and the Hubble constant are also estimated. Furthermore, we find that cosmic expansion is embedded in the gravitational contribution of the frequency-shift within this spacetime metric. Therefore, our results introduce a general relativistic framework that accounts for cosmic expansion and differs from the standard empirical Hubble law.}

\begin{document}
\maketitle
\flushbottom

\section{Introduction}
The existence of a black hole at the center of our galaxy (SgrA*) was proposed after studying the mass distribution in the central region of the Milky Way. The research teams led by Andrea Ghez and Reinhard Genzel  studied the stellar dynamics at the center of our galaxy for decades, independently determining the existence of a supermassive compact object with a mass of the order of $10^6 M_\odot$ \citep{Genzel87, Ghez98, Ghez08, Gillessen09}. 
Another independent observational technique that has provided strong evidence for the existence of supermassive black holes at the very center of active galactic nuclei (AGNs) is the precise measurement of gas dynamics in their accretion disks with the aid of maser-emission lines of water molecule vapor at a wavelength of 1.3 cm, the NGC 4258 spiral galaxy being the most convincing and robust case reported so far with a central supermassive black hole of $4\times10^7 M_\odot$ \citep{Claussen84, Lo2005}. Following this line of research, we are interested in determining the parameters of several black holes by studying the motion of objects within their accretion disks.

Recent works have reported the detection of general relativistic effects from test particles moving in the gravitational field of supermassive black holes. These include the observation of the gravitational redshift \citep{Grav18,Do2019} and the Schwarzschild precession \citep{Grav20} in stars orbiting SgrA$^*$, as well as evidence for the Lense-Thirring effect in M87$^*$ \citep{Iorio25}. The achievement of detecting these general relativistic effects strongly motivates the adoption of relativistic models in the study of supermassive black holes. While the post-Newtonian formalism has yielded satisfactory results, a fully general relativistic approach is likely to provide a more accurate and comprehensive description of these extreme environments.

In \cite{Herrera2015} and \cite{Banerjee2022}, the authors developed a general relativistic model that relates Kerr black hole parameters to astrophysical observables by expressing the redshift and blueshift of photons emitted by a geodesic massive particle in terms of the mass and rotation parameter of the black hole, as well as the radius of the emitting particle. This model has been adapted to a variety of spacetimes, including black hole solutions in general relativity \citep[e.g.][]{Sharif2016, Kraniotis19,Lopez21,Debnath21,Giambo22, Momennia2023,Gerardo}, heterotic string theory \citep{Uniyal17}, quantum gravity \citep{Fu22}, and modified gravity \citep{Sheoran17,Shankar2018,Mustafa22}. Besides, the procedure for obtaining the redshift in any static, spherically symmetric, and asymptotically flat spacetime is detailed in \cite{Diego2024}.

In this work, we consider the Schwarzschild-de Sitter (SdS) metric, also known as the Kottler  background \citep{1918Kottler} with a positive cosmological constant, which describes the spacetime generated by a static, spherically symmetric black hole in a universe dominated by a cosmological constant. Because of the presence of the cosmological constant $\Lambda$ in this metric, the rate of expansion of the universe $H_0$ is naturally introduced in our equations, relating this constant with the black hole mass and distance, photon redshift and the orbital radius of the emitting particle. This relationship enables the estimation of the Hubble constant and black hole parameters from astronomical observational data via Bayesian statistical inference. We perform such an estimation with a model for the frequency shift that incorporates the expansion of the universe through the SdS metric for the first time. The observations required for the statistical fit are the redshift and sky position of objects orbiting the black hole. It is worth mentioning that, for the sake of simplicity, this general relativistic model assumes circular motion on the equatorial plane, conditions satisfied by extragalactic H$_2$O megamaser systems. Within this framework, previous estimations of mass-to-distance ratio, position, and systemic velocity of several megamaser AGN black holes are reported in \cite{Nucamendi21,Villaraos22} and \cite{Adri24} using the Schwarzschild metric  \citep[see][for a comprehensive review of this model]{Adri25}, as well as in \cite{diego25}, by considering conformal gravity for describing the black hole at the core of NGC 4258.

H$_2$O megamasers are composed of water vapor clouds located in the accretion disks of supermassive black holes at the centers of AGNs, emitting at 22 GHz with high luminosity, with the AGN itself as the energy source for stimulated emission \citep{Claussen84, Lo2005}. Numerous observations show that these systems form thin, edge-on disks as seen from the Earth, with three groups of megamasers observed in the sky: the redshifted and blueshifted at the edges of the disk, and the systemic masers located about the line of sight (LOS) \citep{Lo2005}. In addition, their rotation curves demonstrate low eccentricities \citep{Herrnstein99}, confirming that the gas clouds follow nearly circular, equatorial orbits around the central supermassive black hole. This configuration makes H$_2$O megamaser systems optimal astrophysical laboratories for applying the aforementioned general relativistic model of circular geodesic motion.

The Megamaser Cosmology Project (MCP) has conducted most surveys to find extragalactic H$_2$O megamaser systems, discovering 85 of these systems and mapping maser disks in 20 galaxies\footnote{See \url{https://safe.nrao.edu/wiki/bin/view/Main/MegamaserCosmologyProject} for the full catalog.}. These observations are carried out using Very Long Baseline Interferometry (VLBI) with the collaboration of several radio telescopes, measuring the positions and redshifts of the maser features, as well as their acceleration through continued spectral monitoring. VLBI offers the noteworthy submilliarcsecond resolution necessary to resolve the sub-parsec structure of maser disks \citep{Herrnstein99}. The high-precision measurements are publicly available in a series of papers, which also report estimates of the black hole masses and distances obtained by fitting Keplerian rotation curves to the observed maser motion and incorporating relativistic corrections \citep[e.g.][]{MCPIV,MCPVIII,MCPIX, MCPXI, MCPXII}. In addition to characterizing the supermassive black holes in these AGNs, the central objective of the MCP is to estimate the Hubble constant for galaxies at low redshift ($z < 0.05$) and in the Hubble flow ($r_d > 30$ Mpc). Using data from six galaxies, the MCP has reported a value of $H_0 = 73.9 \pm 3.0$ km s$^{-1}$ Mpc$^{-1}$, independent of distance ladders \citep{MCPXIII}.

The discrepancy between local and early-universe measurements of the Hubble constant remains unsolved despite consideration of systematic errors and improvement in observational precision. Several studies have explored the Hubble constant within the standard Friedmann-Lemaitre-Robertson-Walker (FLRW) cosmological framework, using different classes of astrophysical probes \citep{Dainotti25}. Local measurements based on observations of Cepheids and Type Ia Supernovae with the Hubble Space Telescope yield a value of $H_0 = 73.30 \pm  1.04$ km s$^{-1}$ Mpc$^{-1}$ \citep{Riess22}, a result confirmed by recent observations from the James Webb Space Telescope \citep{Riess24}. In contrast, early-universe measurements based on the cosmic microwave background yield a value of $H_0 = 67.4 \pm 0.5$ km s$^{-1}$ Mpc$^{-1}$ \citep{Planck20}. Thus, using alternative independent methods to estimate the Hubble constant represents great interest and importance, like the recent fit provided in \cite{Freedman24} and \cite{ Lee24} using J-region asymptotic giant branch observations from the James Webb Space Telescope. In a complementary context, constraints on the cosmological constant have also been obtained from the dynamics of local gravitational systems, such as galaxy rotation curves \citep{Benisty24}. 

Motivated by investigating this tension further, our primary goal is to estimate the Hubble constant using the aforementioned general relativistic formalism and megamaser observations, which are independent of traditional distance indicators \citep{MCPXIII}. Additionally, we estimate the mass, distance, and position of five megamaser AGN black holes, modeled by the SdS metric. This background serves as a first approximation to explore the gravitational effects caused by black holes and the cosmological constant, while avoiding the complexities introduced by matter in the universe. This initial approach is useful for analyzing the relationship between black holes and cosmological properties, and can be extended in future research to more realistic models that include matter and additional cosmological factors. 

This paper it divided as follows. In section \ref{sec:2} we obtain the expressions for the frequency-shift of photons within the formalism of general relativity, explicitly showing the relation between black hole properties, the cosmological constant, and the redshift. In section \ref{sec:stats}, we apply the theoretical model to megamaser systems, incorporating corrections for small deviations from the idealized model. Then, we perform a Bayesian statistical fit using the Markov Chain Monte Carlo (MCMC) method to estimate the parameters. The estimations are carried out individually for each system, followed by a simultaneous estimation using the data from the five galaxies. In section \ref{sec:results} we present the results obtained for the Hubble constant and the black hole parameters from these fits. Finally, in section \ref{sec:conclusion} we discuss our results.

\section{Redshift in Schwarzschild-de Sitter (Kottler) spacetime}
\label{sec:2}
For the Kerr-de Sitter metric, the authors in \cite{Momennia2023} provide a comprehensive derivation for the frequency shift of photons in terms of spacetime parameters: mass, spin, and cosmological constant, and particularly in the language of the Hubble constant. Due to the negligible effect of the black hole spin on the dynamics of the megamaser systems, since their orbits are located at hundreds of thousands of Schwarzschild radii, in this section, we set the rotation parameter $a = 0$ and present the corresponding relations for the SdS background.

The black hole solution to the Einstein equations 
with cosmological constant is the SdS metric. In geometrized units, its line element reads
\begin{align}
    ds^2 = - \left( 1- \frac{2M}{r}- \frac{\Lambda r^2}{3} \right) dt^2 + \left( 1- \frac{2M}{r}- \frac{\Lambda r^2}{3} \right) ^{-1} dr^2 + r^2 d\Omega ^2 ,
\end{align}
where $M$ is the black hole mass, $\Lambda$ the cosmological constant, and the line element of the unitary two-sphere reads $d\Omega^2 = d\theta^2 + \sin^2\theta d\phi^2$.

This metric, in this coordinate patch, describes a spacetime with an event horizon $r_{EH}$ generated by the black hole mass and a cosmological horizon $r_{CH}$, with $\Lambda$ as the source of the universe's expansion. 

\begin{figure}
        \centering
        \includegraphics[width=0.8\linewidth]{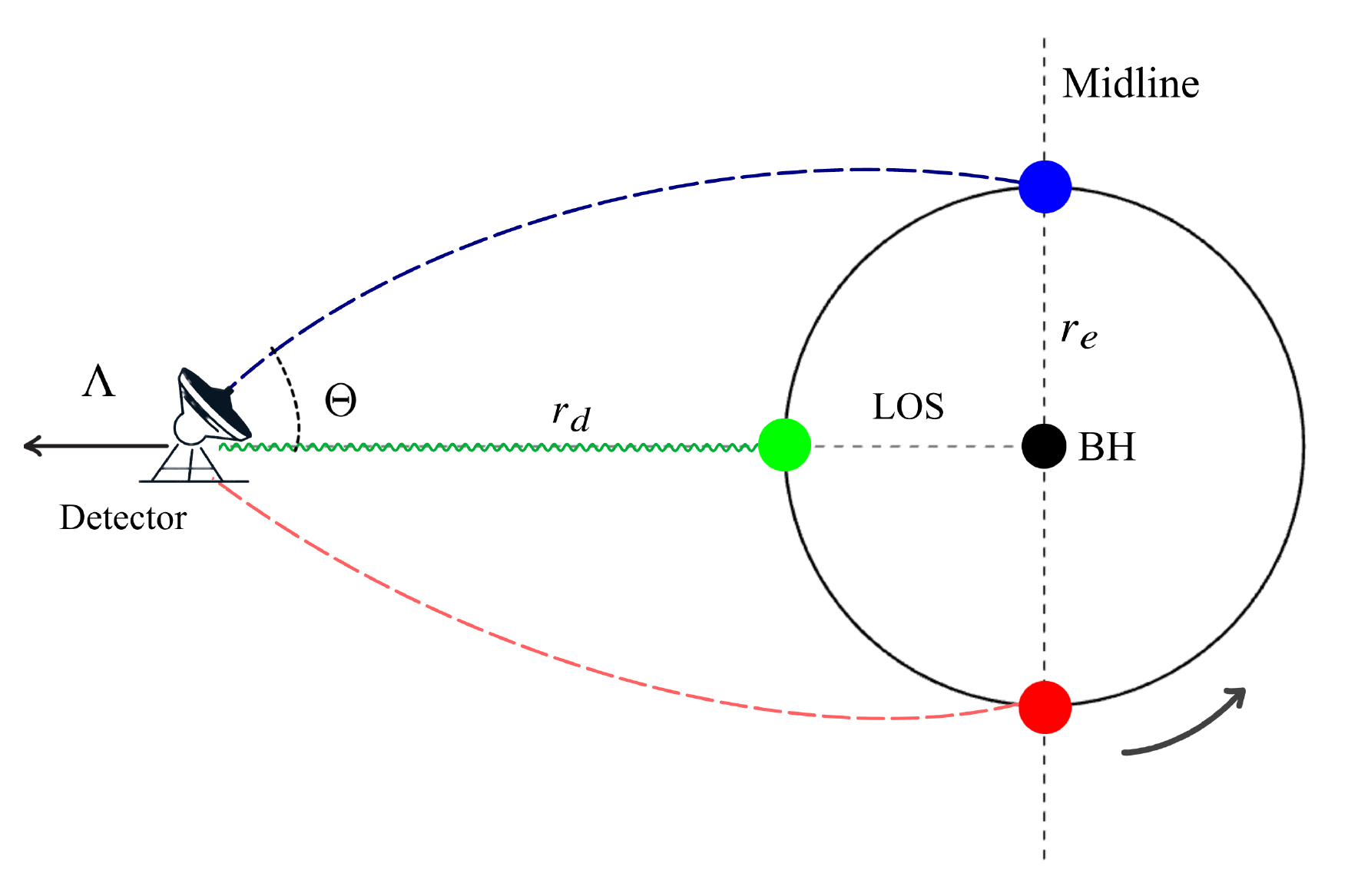}
        \caption{A photon source at $r_e$ circularly orbiting the SdS black hole and an observer  under the action of the cosmological constant $\Lambda$ radially receding away from the black hole. This observer is located along the LOS at a distance $r_d$, where it measures the frequency shift ($Z_{obs}$) and the position on the sky ($\Theta$) of the photon source. Megamaser observations show three groups of maser features on the sky: those about the midline, and the ones along the LOS.}
        \label{fig:model}
\end{figure}

We now consider a photon source consisting of a massive particle circularly orbiting this SdS black hole. We aim to derive the expression of the frequency-shift as detected by a distant observer in terms of spacetime parameters, i.e., the mass of the black hole, the cosmological constant, and the radii of the emitter and the detector. See figure \ref{fig:model} for a representation of this system.

\subsection{Geodesic motion of massive particles}

We start by considering a massive particle in circular geodesic motion  with radius $r_e$ on the equatorial plane around the SdS black hole, with four velocity $U^\mu$. This metric has timelike $\xi^\mu$ and azimuthal $\psi^\mu$ Killing vectors associated with the conserved quantities of energy $E$ and angular momentum $L_\phi$,
\begin{align}
    U^{\mu} \xi_\mu & = - E , & U^{\mu} \psi_\mu & = L_\phi .
\end{align}
By using the normalization of the four-velocity $U^\mu U_\mu = -1$, we obtain expressions for $E$ and $L_\phi$, and therefore, the non-zero components of the four-velocity in terms of the spacetime parameters
\begin{align}
    U_e^t (r_e, \pi /2) & = \frac{1}{\sqrt{1- \dfrac{3M}{r_e}}}  ,
    \label{eq:U^t_e}
\end{align}
\begin{align}
     U_e^{\phi} (r_e, \pi /2)& = \pm  \dfrac{1}{r_e} \sqrt{\dfrac{\dfrac{M}{r_e} - \dfrac{\Lambda r_e^2}{3}}{1-\dfrac{3M}{r_e}}} ,
     \label{eq:U^phi_e}
\end{align}

\subsection{Geodesic motion of photons}
Photons, on the other hand, travel in null geodesics from the emission point to a far-away observer, with four-momentum $k^\mu$ such that $k^\mu k_\mu =0$. Similar to the massive case, the conserved quantities associated to the Killing vectors are the photon's  energy $E_\gamma$ and angular momentum $L_\gamma$, leading to the following expressions 
\begin{align}
    k^{t}& = - \frac{E_\gamma}{g_{tt}} , & k^{\phi} & =  \frac{L_\gamma}{g_{\phi\phi}} ,
\label{eq:ktp}
\end{align}
\begin{equation}
    \left( \frac{k^r }{E_\gamma} \right)^2= 1 - \frac{L_\gamma^2 }{E_\gamma^2 r^2}\left( 1- \frac{2M}{r}   -\frac{\Lambda r^2}{3} \right).
    \label{eq:k_r_E_L}
\end{equation}

The light bending parameter or apparent impact factor is defined as the conserved quantity $b = L_\gamma/E_\gamma$. At the extremes of the orbit ($\phi \approx \pm \pi/2$), the massive particle has its maximum tangential velocity relative to the observer, maximizing the chances of detecting the photons'  redshift and blueshift. We focus on these points where the light bending parameter is given by
\begin{equation}
    b_{e,\mp} = \mp \dfrac{r_e}{\sqrt{1- \dfrac{2M}{r_e}   - \dfrac{\Lambda r_e^2}{3} }} . 
    \label{eq:b}
\end{equation}

\subsection{Geodesic motion of a distant detector}

Now, we consider a distant detector in the equatorial plane, at a distance $r_d$ from the black hole. This detector is located between the zero gravity radius (ZGR), at which the gravitational influences of the black hole and the cosmological constant cancel each other out, and the cosmological horizon. Considering only radial motion due to the expansion of the universe, the non-zero components of the detector's four-velocity are
\begin{align}
    U_d^t(r_d, \pi/2) = \frac{\sqrt{1-(9M^2\Lambda)^{1/3}}}{1- \dfrac{2M}{r_d} - \dfrac{\Lambda r_d^2}{3} },
    \label{eq:U^t_d}
\end{align}
\begin{align}
    U_d^r(r_d, \pi/2) = \sqrt{\frac{2M}{r_d} + \frac{\Lambda r_d^2}{3} - (9M^2\Lambda)^{1/3}},
    \label{eq:U^r_d}
\end{align}

\subsection{Frequency-shift}

The redshift ($Z_{SdS{+}}$) and blueshift ($Z_{SdS{-}}$) of the photons are defined, in terms of the frequency $\omega$, by 
\begin{align}
    1 + Z_{SdS {\pm}} = \frac{\omega_e}{\omega_d} = \frac{-(k_\mu U ^\mu)|_e}{-(k_\mu U ^\mu)|_d} ,
    \label{eq:zsds_kU}
\end{align}
where the subscripts $_e$ and $_d$ refer to the emission and detection points, respectively. 

Then, we substitute equations \eqref{eq:U^t_e}-\eqref{eq:U^r_d} into \eqref{eq:zsds_kU} to obtain the full expression of the redshift
\begin{equation}
    1 + Z_{SdS{\pm}} = \dfrac{1}{\sqrt{1-3\tilde{M}}} 
    \frac{\left(1 \pm \sqrt{\dfrac{\tilde{M} - \tilde{\Lambda}}{1- 2\tilde{M} - \tilde{\Lambda} }}\right) \left( 1 - 2\bar{M} - \bar{\Lambda} \right)}{ \sqrt{1-(9M^2\Lambda)^{1/3}}  - \sqrt{2\bar{M} + \bar{\Lambda} - (9M^2\Lambda)^{1/3} } \sqrt{1 - \frac{r_e^2}{r_d^2} \frac{1 - 2\bar{M} - \bar{\Lambda} }{1 - 2\tilde{M} - \tilde{\Lambda} }  } } ,
\label{eq:fullredshift}
\end{equation}
where we redefined the variables as $\tilde{M}= M/r_e$, $\tilde{\Lambda} = \Lambda r_e^2/3 $, $\bar{M} = M/r_d$, and $\bar{\Lambda} =  \Lambda r_d^2/3$. This redshift arises from three contributions: the mass of the black hole, the azimuthal motion of the particle, and the expansion of the universe.

In general, the redshift \eqref{eq:zsds_kU} can be expressed as the sum of the temporal and the azimuthal contributions
\begin{equation}
    Z_{SdS{\pm}} = Z_g + Z_{kin_{\pm}},
\end{equation}
where $Z_g$ is the temporal contribution associated with the gravitational redshift, arising from the curvature of spacetime caused by the black hole mass
\begin{equation}
    1 + Z_{g} = \dfrac{U^t_e}{U^t_d - \dfrac{k_r}{E_\gamma}U^r_d}.
    \label{eq:zgrav}
\end{equation}

Conversely, the azimuthal contribution corresponds to the kinematic redshift caused by the motion of the particle within the black hole's gravitational field
\begin{equation}
    Z_{kin_{\pm}} = -\dfrac{b_{e,\mp }U^\phi_e}{U^t_d - \dfrac{k_r}{E_\gamma}U^r_d}.
    \label{eq:zkin}
\end{equation}

The redshift due to the expansion of the universe is embedded in both the gravitational and the kinematic redshifts because of the presence of the cosmological constant in these expressions.

Considering that the cosmological constant is of the order of $\Lambda \sim 10^{-52}$ m$^{-2}$, that the detector is at a considerable distance from the source ($r_d>30$ Mpc assuming that the black hole is within the Hubble flow), and that the radius of the emitter is at sub-parsec scale ($r_e<1$ pc), we expand equation \eqref{eq:fullredshift} for $M/r_d \rightarrow 0$, $\Lambda r_e^2 \rightarrow 0$ and $\Lambda r_d^2 \rightarrow 0$. Keeping the first dominant term of $\Lambda r_d^2$, the SdS redshift is expressed as the following product \citep{Momennia2023}
\begin{equation}
    1 + Z_{SdS{\pm}} = (1 + Z_{Schw\pm})(1+Z_\Lambda),  
    \label{eq:z_schw_L}
\end{equation}
with $Z_{Schw\pm}$ as the frequency shift in the Schwarzschild spacetime, reading \citep[see][]{Nucamendi21}
\begin{equation}
    Z_{Schw{\pm} } = \sqrt{\frac{1}{1-3\tilde{M}}} \pm \sqrt{\frac{\tilde{M}}{(1-3\tilde{M})(1-2\tilde{M})}},
\end{equation}
and $Z_\Lambda$ represents the cosmological redshift,
\begin{equation}
Z_{\Lambda} = \sqrt{\frac{\Lambda}{3}} r_d = H_0 r_d.
\label{eq:zlambda}
\end{equation}
Here we used the relation between the Hubble constant and the cosmological constant for a de Sitter universe 
\begin{equation}
    H_0 = \sqrt{\frac{\Lambda}{3}}.
    \label{eq:hub_Lambda}
\end{equation}
Note that equation \eqref{eq:zlambda} is the Hubble law.  

Hence, a general relativistic version of the Hubble law is contained in equation \eqref{eq:fullredshift} as the cosmological redshift for a particle in circular motion around a black hole receding away from us. Unlike other metrics with no cosmological constant, in which we have to introduce the recessional redshift manually, the cosmological redshift arises naturally in the SdS spacetime.

\subsection{Redshift in terms of the Hubble constant and astrophysical observables}

We derived the full expression of the redshift \eqref{eq:fullredshift} in terms of the mass, the cosmological constant and the radii of emission and detection. Nevertheless, the radius of emission is not an observable; the related quantity that we observe is the position on the sky of the emitting source $\Theta$. From figure \ref{fig:model}, it is clear that the relation between the radius of emission at the midline and the observed sky position for large distances is $r_e \approx \Theta r_d$. Thus, we express equation \eqref{eq:fullredshift} in terms of the black hole mass, the radius of detection (or distance to the black hole), the measured angular position of the source, and the Hubble constant
\begin{equation}
    1 + Z_{SdS{\pm}} = \dfrac{1}{\sqrt{1-3\frac{\bar{M}}{\Theta}}} 
    \frac{\left(1 \pm \sqrt{\dfrac{\frac{\bar{M}}{\Theta} - \bar{H_0}^2\Theta^2}{1- 2\frac{\bar{M}}{\Theta} - \bar{H_0}^2\Theta^2 }}\right) \left( 1 - 2\bar{M} - \bar{H_0}^2 \right )}{ \sqrt{1-3(MH_0)^{2/3}}  - \sqrt{2\bar{M} + \bar{H_0}^2 - 3(MH_0)^{2/3} } \sqrt{1 - \frac{\Theta^2 (1 - 2\bar{M} - \bar{H_0}^2 )}{1 - 2\frac{\bar{M}}{\Theta} - \bar{H_0}^2\Theta^2 }  } } ,
\label{eq:Z_H0_Theta}
\end{equation}
with $\bar{H_0} = H_0r_d$. Note that the mass, distance, and the Hubble constant are coupled given the presence of the products $M/r_d$ and $H_0r_d$.
\begin{figure*}
\centering
\begin{subfigure}{0.478\linewidth}
\includegraphics[width=\linewidth]{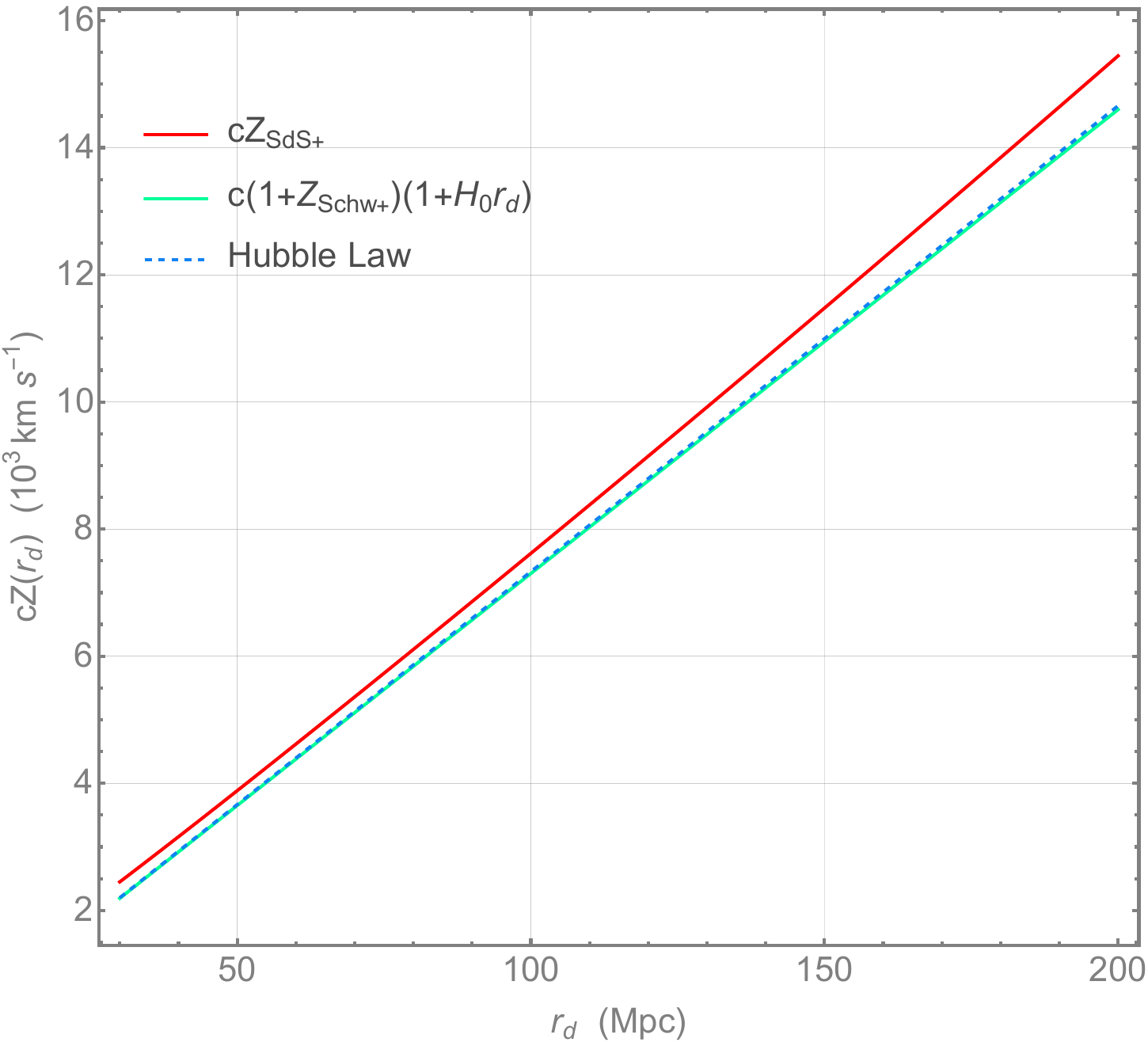}
\caption{Redshifts}
\label{fig:redshifts}
\end{subfigure}
\begin{subfigure}{0.48\linewidth}
\includegraphics[width=\linewidth]{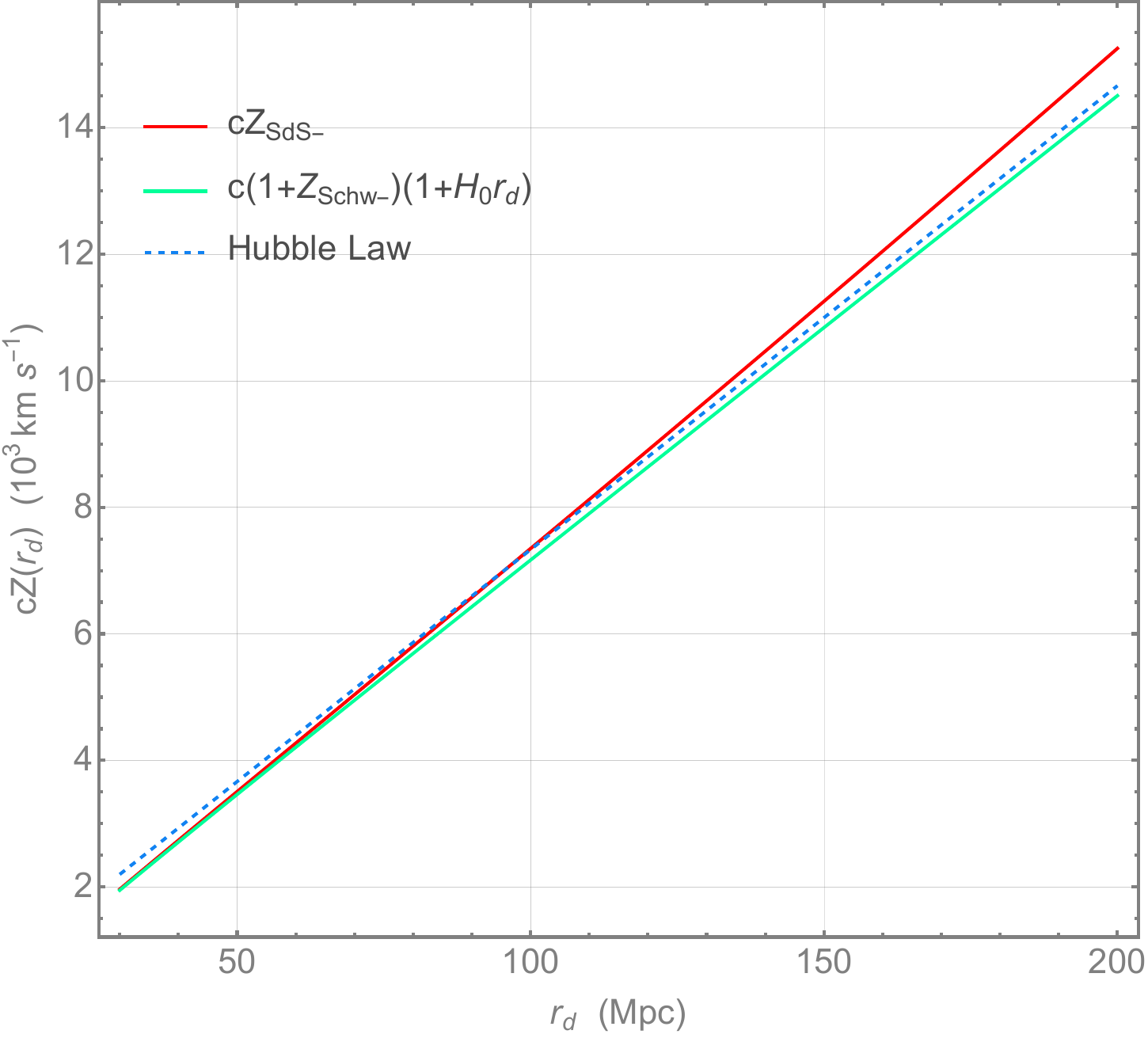}
\caption{Blueshifts}
\label{fig:blueshifts}
\end{subfigure}
\caption{Plots of the redshift (a) and blueshift (b) for SdS (red), for Schwarzschild composed with the cosmological redshift (green), and for the Hubble law (dotted line).}
\label{fig:redblue}
\end{figure*}

In figure \ref{fig:redblue}, we plot the redshift and blueshift, expressed as velocities, as functions of distance for different cases: The redshift in the SdS spacetime  \eqref{eq:Z_H0_Theta}, the composed redshift of the Schwarzschild metric with the cosmological one that follows the Hubble law \eqref{eq:z_schw_L}, 
and the redshift corresponding to the Hubble law alone \eqref{eq:zlambda}. For this plot, we adopt a black hole mass of $M = 1 \times 10^7M_\odot$, an angular distance of the source of $\Theta = 0.3$ mas, and a Hubble constant $H_0 = 73.3$ km s$^{-1}$ Mpc$^{-1}$, with the detector distance ranging between 30 and 200 Mpc, a distance range that comprises all of our galaxies. The values of mass, angular distance of the source, and distance of the detector were chosen to be consistent with typical megamaser systems hosting supermassive black holes, while the value of $H_0$ follows ref. \cite{Riess22}.

From figure \ref{fig:redshifts} we observe that the first two cases exhibit redshifts larger than the Hubble law. Conversely, in figure \ref{fig:blueshifts}, the corresponding blueshifts are smaller than those predicted by the Hubble law for distances between 30 Mpc and 100 Mpc. This difference originates from the contribution of the kinematic redshift, whose effect is more evident at small distances and becomes clearly noticeable for $r_d < 80$ Mpc. At distances larger than 100 Mpc, where the expansion effect dominates the frequency shift, the SdS redshift yields the largest redshift. The case of the composition of the Schwarzschild and the cosmological redshifts is the one that most closely resembles the Hubble law, showing a linear behavior after 100 Mpc.

The difference between the frequency-shift predicted by the pure Hubble law and that obtained in the SdS spacetime arises from the inclusion of the black hole mass and the azimuthal motion of the emitting particle. In the Hubble law, the only quantity that influences the redshift (for fixed $H_0$) is the distance to the source, whereas in the SdS case the redshift additionally encodes both gravitational and kinematic contributions. To isolate the role of gravity alone, in appendix \ref{app:Static} we examine the case of a static photon source, thereby removing any kinematic redshift and highlighting the gravitational effects of the black hole mass and the cosmological constant.

\section{Bayesian analysis applied to megamaser systems}
\label{sec:stats}
\subsection{Application to megamaser data}
Here, we apply the model presented in the previous section to megamaser systems consisting of water vapor clouds. These systems, which are observed nearly edge-on, fit Keplerian rotation curves consistent with nearly circular motion \citep{mcpii,mcpvii}.
The MCP provides measurements of individual maser features using VLBI techniques. These measurements include the angular position in mas $(\Theta_x,\Theta_y)$ with associated uncertainties $(\sigma_{\Theta_x},\sigma_{\Theta_y})$. The angular distance of the maser features to the black hole is thus   $\Theta = \sqrt{(\Theta_x - x_0)^2+(\Theta_y - y_0)^2}$, where $(x_0, y_0)$ is the position of the black hole. In addition, the observed redshift $Z_{obs}$ is reported as a velocity relative to the local standard of rest by adopting the optical definition of  redshift $v = cZ$. 

Since we aim to constrain the Hubble constant, we focus on the galaxies located at greater distances, where the influence of peculiar velocities becomes comparatively small, thus providing more reliable constraints on the cosmic expansion rate. Additionally, to obtain high-precision results, we choose galaxies with a large number of data, whose measurements are among the most precise. Based on these criteria, the selected galaxies from the MCP catalog are UGC 3789 \citep{MCPIV}, NGC 5765b \citep{MCPVIII}, NGC 6264 \citep{mcpv}, NGC 6323 \citep{mcpvi}, and CGCG 074-064 \citep{MCPXI}.

\subsection{Bayesian modeling}
\label{sec:bayesmod}
For each megamaser system, the set of free parameters is $\Omega=(M/r_d, H_0r_d, x_0, y_0)$. We constrain these parameters by maximizing the likelihood $\mathcal{L}$ defined as
\begin{align}
\mathcal{L} = \sum_i \frac{1}{\sqrt{2\pi}\sigma_i}\exp\left[ -\frac{1}{2}\dfrac{(Z_{obs,i} - Z_{mod,i}(\Omega))^2}{\sigma_i^2} \right],
\label{eq:likelihood}
\end{align}
where $Z_{obs,i}$ is the i-th observed data, which in our case, is the redshift data for each megamaser, $Z_{mod,i}$ is the expected value of the redshift from the theoretical model given by equation \eqref{eq:Z_H0_Theta}, and $\sigma$ is the observational error.

We measure the goodness of fit via the Pearson's $\chi^2$, defined by its relation to the likelihood $\ln{\mathcal{L}} \propto -\frac{1}2{\chi^2}$, hence $\chi^2$ reads
\begin{align}
    \chi^2 = \sum_{i=1}{ \dfrac{\left[Z_{obs,i} - (Z_{g} + \cos{(\delta \phi)} \sin{\theta_0} Z_{kin_{\pm}})\right]^2}{\delta Z_{SdS_{\pm}}^2 + \left( \delta\phi\sin{(\delta\phi)} Z_{kin_{\pm}} \right)^2 }}.
    \label{eq:chi_redshift}
\end{align}

In the model, we further considered that the maser disk slightly deviates from the equatorial plane such that the inclination angle $\theta_0 \approx \pi /2$, and that the masers are spread about the midline by a small scattering angle $\delta \phi$, such that $\phi = \pi/2 \pm \delta\phi$. The term $\delta Z_{SdS{\pm}}$ is the uncertainty on the redshift
\begin{align}
    \delta Z_{SdS_{\pm}} = \delta Z_{g} \pm \cos{\delta \phi} \sin \theta_0\delta Z_{kin_+}.
\end{align}

The expressions for the gravitational and kinematic redshift uncertainties, $\delta Z_{g}$ and $\delta Z_{kin_{\pm}}$, respectively, are derived to propagate measurement errors in the angular position into the redshift estimates

\begin{equation}
\begin{split}
    & \delta Z_{g} = - \frac{1+Z_{g}}{1-3\frac{\bar{M}}{\Theta}} \left(\frac{\delta \Theta}{\Theta} \right) \left\{ \frac{3}{2} \frac{\bar{M}}{\Theta } + \frac{\Theta^2 \left(1+Z_{g}\right)\left(1-3\frac{\bar{M}}{\Theta}\right)^{3/2}}{1 - 2\frac{\bar{M}}{\Theta} - \bar{H_0}^2 \Theta^2} \right. \times 
    \\
& \left. \left[  \dfrac{\left(2\bar{M} + \bar{H_0}^2 - 3(MH_0)^{2/3}\right)\left(1 - 2\frac{\bar{M}}{\Theta} - \bar{H_0}^2\Theta^2\right)}{1 - \frac{2\bar{M}}{\Theta} -\Theta^2\left(1 - 2\bar{M}\right)} + \left(\frac{\bar{M}}{\Theta} -\bar{H_0}^2\Theta^2   \right ) \right]  \right\} ,
\end{split}
\label{eq:dzgrav}
\end{equation}
\begin{equation}
      \delta Z_{kin{\pm}} =  \delta Z_{g} \sqrt{\dfrac{\frac{\bar{M}}{\Theta}-\bar{H_0}^2\Theta^2}{1-2\frac{\bar{M}}{\Theta}-\bar{H_0}^2\Theta^2}} - Z_{kin{\pm}} \left(\frac{\delta \Theta}{\Theta} \right) \left[ \dfrac{\frac{\bar{M}}{2\Theta} + \bar{H_0}^2 \Theta^2}{\frac{\bar{M}}{\Theta}- \bar{H_0}^2 \Theta^2} + \dfrac{ \frac{\bar{M}}{\Theta}-\bar{H_0}^2 \Theta^2}{1- 2 \frac{\bar{M}}{\Theta}- \bar{H_0}^2 \Theta^2}\right],
  \label{eq:dz}
\end{equation}
where $\delta\Theta$ is the variation on the angular position, which depends on the observational uncertainties and is given by
\begin{align}
    \delta \Theta = \sqrt{\left(\frac{\Theta_x - x_0}{\Theta}\right)^2 \sigma_{\Theta_x}^2 + \left(\frac{\Theta_y - y_0}{\Theta}\right)^2 \sigma_{\Theta_y}^2 }.
\end{align}

\section{Results}
\label{sec:results}

We perform a Bayesian fit using the MCMC method, implemented with the \texttt{emcee} Python package \citep{emcee}. Following the likelihood formulation in section \ref{sec:bayesmod}, the MCMC samples the posterior distribution of the parameters using the observed redshifts and their uncertainties. The observational data for each source are taken from  \cite{MCPIV, mcpv, mcpvi, MCPVIII} and \cite{MCPXI}, from where we also retrieve the inclination angles $\theta_0$ needed in equation \eqref{eq:chi_redshift}. The scattering angle $\delta\phi$ for each galaxy is set so that the reduced Pearson's $\chi^2$ approaches unity. In addition, we move the original coordinate origin of the maser features to the geometric center formed by the LOS masers and rotate them so that the maser disk lies horizontally on the equatorial plane, as done in ref. \cite{Villaraos22}.

The free parameters $\Omega = (M/r_d, H_0r_d, x_0, y_0)$ are independently constrained for each system, adopting uniform priors for all of them. We also carry out a joint analysis by combining the five galaxies to obtain a single constraint on the Hubble constant. In this case, we impose a Gaussian prior on $H_0$ based on the result in \cite{Riess22}, $H_0 = 73.30 \pm  1.04$ km s$^{-1}$ Mpc$^{-1}$. This choice is motivated by the fact that the galaxies lie within $r_d<150$ Mpc, corresponding to the late Universe. Given the degeneracy between the mass, distance and the Hubble constant, setting this Gaussian prior allows us to estimate these quantities separately.

\begin{table*}[]
\centering
\resizebox{\textwidth}{!}{
\begin{tabular}{ c c c c c c c }
\hline
Source & $M/r_d$ & $H_0r_d$ & $x_0$ & $y_0$ & $\delta\phi$&$\chi^2/N$ \\
& ($10^5 M_{\odot}$ Mpc$^{-1}$) &  (km s$^{-1}$) & (mas) & (mas) & ($\degree$) &\\
\hline \hline
\multirow{2}{*}{UGC 3789} & \multirow{2}{*}{$2.242 \pm0.009$} & \multirow{2}{*}{$3281.00 \pm 2.89$} & \multirow{2}{*}{$-0.049 \pm 0.006$} & \multirow{2}{*}{$0.011 \pm 0.025$} & \multirow{2}{*}{7} & \multirow{2}{*}{1.47} \\
& & & &  &\\
\hline
\multirow{2}{*}{NGC 5765b} & \multirow{2}{*}{$3.670 \pm 0.013$} & \multirow{2}{*}{$8314.62 \pm 2.36$} & \multirow{2}{*}{$0.122 \pm 0.008$} & \multirow{2}{*}{$-0.100\pm 0.034$} & \multirow{2}{*}{$7$} & \multirow{2}{*}{1.18} \\
& & & &  &\\
\hline 
\multirow{2}{*}{NGC 6264} & \multirow{2}{*}{$2.093\pm 0.007$} & \multirow{2}{*}{$10212.32 \pm 3.44$} & \multirow{2}{*}{$0.005 \pm 0.006$} & \multirow{2}{*}{\ldots} & \multirow{2}{*}{7} & \multirow{2}{*}{1.33}  \\
& & & & &\\
\hline 
\multirow{2}{*}{NGC 6323} & \multirow{2}{*}{$0.901 \pm 0.005$} & \multirow{2}{*}{$7840.87 \pm 3.93$} & \multirow{2}{*}{$0.018 \pm 0.007$} & \multirow{2}{*}{$0.012 \pm 0.029$} & \multirow{2}{*}{8} & \multirow{2}{*}{1.20}  \\
& & & &  &\\
\hline 
\multirow{2}{*}{CGCG 074-064} & \multirow{2}{*}{$2.704 \pm 0.018$} & \multirow{2}{*}{$6947.07 \pm 5.65$} & \multirow{2}{*}{$-0.041 \pm 0.007$} & \multirow{2}{*}{$0.044 \pm 0.020$} & \multirow{2}{*}{8} &  \multirow{2}{*}{1.20} \\
& & & &  &\\
\hline 
\end{tabular} }
\caption{Posterior values of the individual fits for each source. Namely, the mass-to-distance ratio, the product of the Hubble constant with the distance, and the black hole offset. For NGC 6264 the $y_0$ offset is fixed at the geometric center of the disk formed by the high-redshifted maser features. The best scattering angle at which the maser features
are spread on the azimuthal angle about the midline and the reduced $\chi^2$ of the fit are also shown. }
\label{tab:ind_res}
\end{table*}

\begin{table*}
\centering
\resizebox{\textwidth}{!}{
\begin{tabular}{ c c c c c c c }
\hline
Source & $M$ & $r_d$ & $x_0$ & $y_0$ & $H_0$ & $\chi^2/N$ \\
& ($10^7 M_{\odot}$) &  (Mpc) & (mas) & (mas) &  (km s$^{-1}$ Mpc$^{-1}$) \\
\hline   \hline
\multirow{2}{*}{UGC 3789} & \multirow{2}{*}{ $1.010\pm0.015$} & \multirow{2}{*}{$45.01^{+0.64}_{-0.62}$ } & \multirow{2}{*}{$-0.049 \pm 0.006$} & \multirow{2}{*}{$0.011 ^{+0.024}_{-0.025}$} & \multirow{10}{*}{$72.89\pm 1.02$} &\multirow{10}{*}{1.27} \\
& & & & & \\
\cline{1-5}
\multirow{2}{*}{NGC 5765b} & \multirow{2}{*}{$4.186 \pm 0.060$ } & \multirow{2}{*}{$114.07  ^{+1.62}_{-1.58}$} & \multirow{2}{*}{$0.123 \pm 0.008$} & \multirow{2}{*}{$-0.100 \pm 0.033$} \\
& & & & & \\
\cline{1-5}
\multirow{2}{*}{NGC 6264} & \multirow{2}{*}{$2.932^{+0.043}_{-0.042} $} & \multirow{2}{*}{$140.11 ^{+1.99}_{-1.94}$} & \multirow{2}{*}{$0.005 \pm 0.006$} & \multirow{2}{*}{\ldots} & &\\
& & & & & \\
\cline{1-5}
\multirow{2}{*}{NGC 6323} & \multirow{2}{*}{$0.970 \pm 0.015$} & \multirow{2}{*}{$107.57^{+1.53}_{-0.04}$} & \multirow{2}{*}{$0.018 ^{+0.007}_{-0.028}$} & \multirow{2}{*}{$0.013 ^{+0.029}_{-1.492}$} & & \\
& & & & & \\
\cline{1-5}
\multirow{2}{*}{CGCG 074-064} & \multirow{2}{*}{$2.578 \pm 0.040$} & \multirow{2}{*}{$95.31  ^{+1.35}_{-1.32}$} & \multirow{2}{*}{$-0.041 \pm 0.007$} & \multirow{2}{*}{$0.044\pm 0.020$}  & &\\
& & & & & \\
\hline  
\end{tabular} }
\caption{Posterior values of the joint fit where a Gaussian prior on $H_0$ was used. The estimated parameters are the mass, distance, and position of the black hole, as well as the Hubble constant and the reduced $\chi^2$ of this fit.}
\label{tab:joint_res}
\end{table*}

The posterior individual estimations of the free parameters of mass-to-distance ratio, the product of the Hubble constant and the black hole distance from Earth, and the black hole position for the galaxies UGC 3789, NGC 5765b, NGC 6264, NGC 6323, and CGCG 074-064 are presented in table \ref{tab:ind_res}. This table also lists the best scattering angles, as well as the values of the reduced $\chi^2$. The results of the joint fit, where the data from the five systems are analyzed simultaneously, are summarized in table \ref{tab:joint_res}. For this fit, we estimate separately the quantities of mass, distance, angular position of the black hole, and the Hubble constant, given the use of a Gaussian prior.

\begin{figure*}[]
\centering
\begin{subfigure}{0.3\linewidth}
\includegraphics[width=\linewidth]{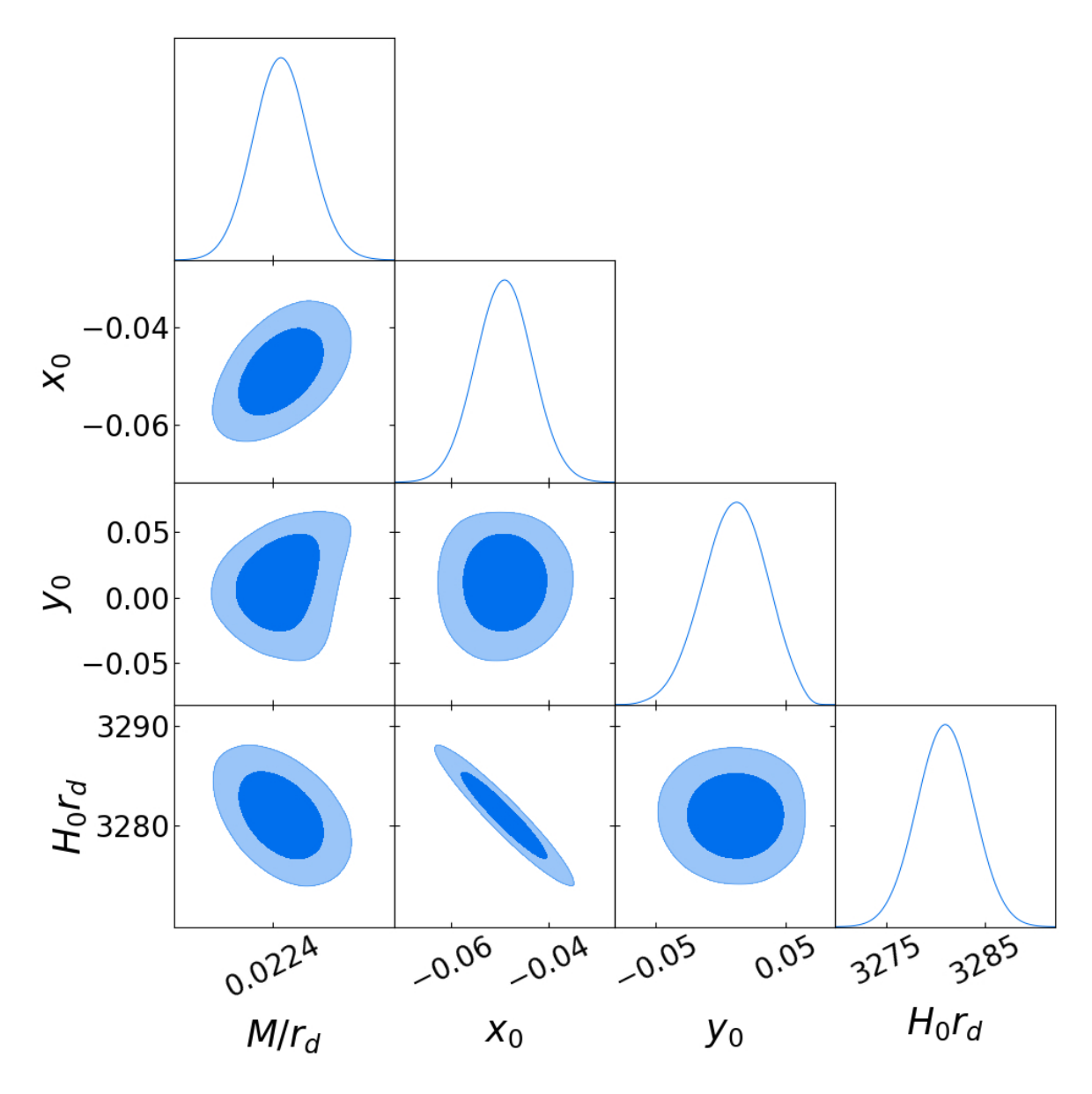}
\caption{UGC 3789}
\label{fig:3789}
\end{subfigure}
\begin{subfigure}{0.3\linewidth}
\includegraphics[width=\linewidth]{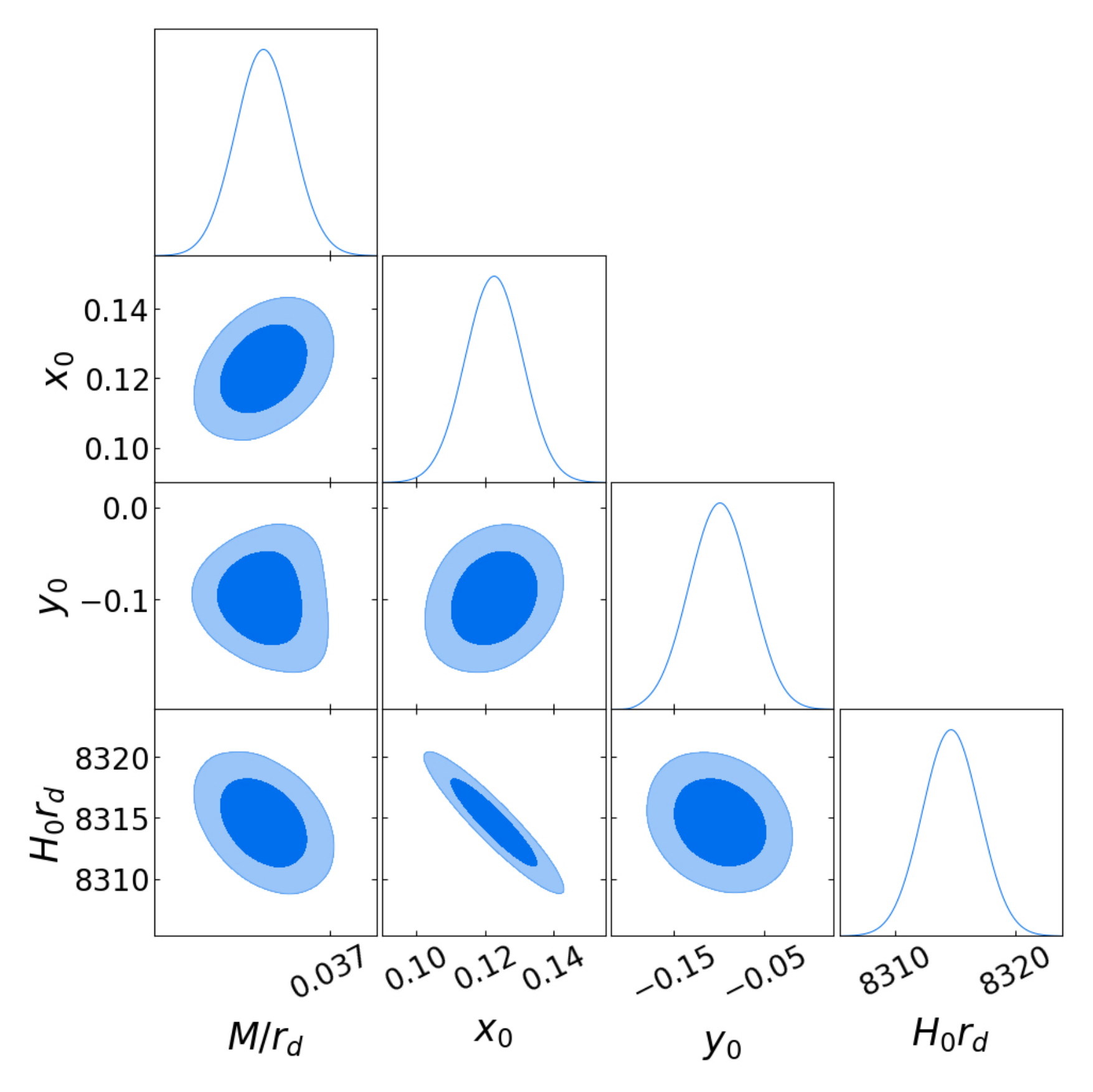}
\caption{NGC 5765b}
\label{fig:5765b}
\end{subfigure}
\begin{subfigure}{0.3\linewidth}
\includegraphics[width=\linewidth]{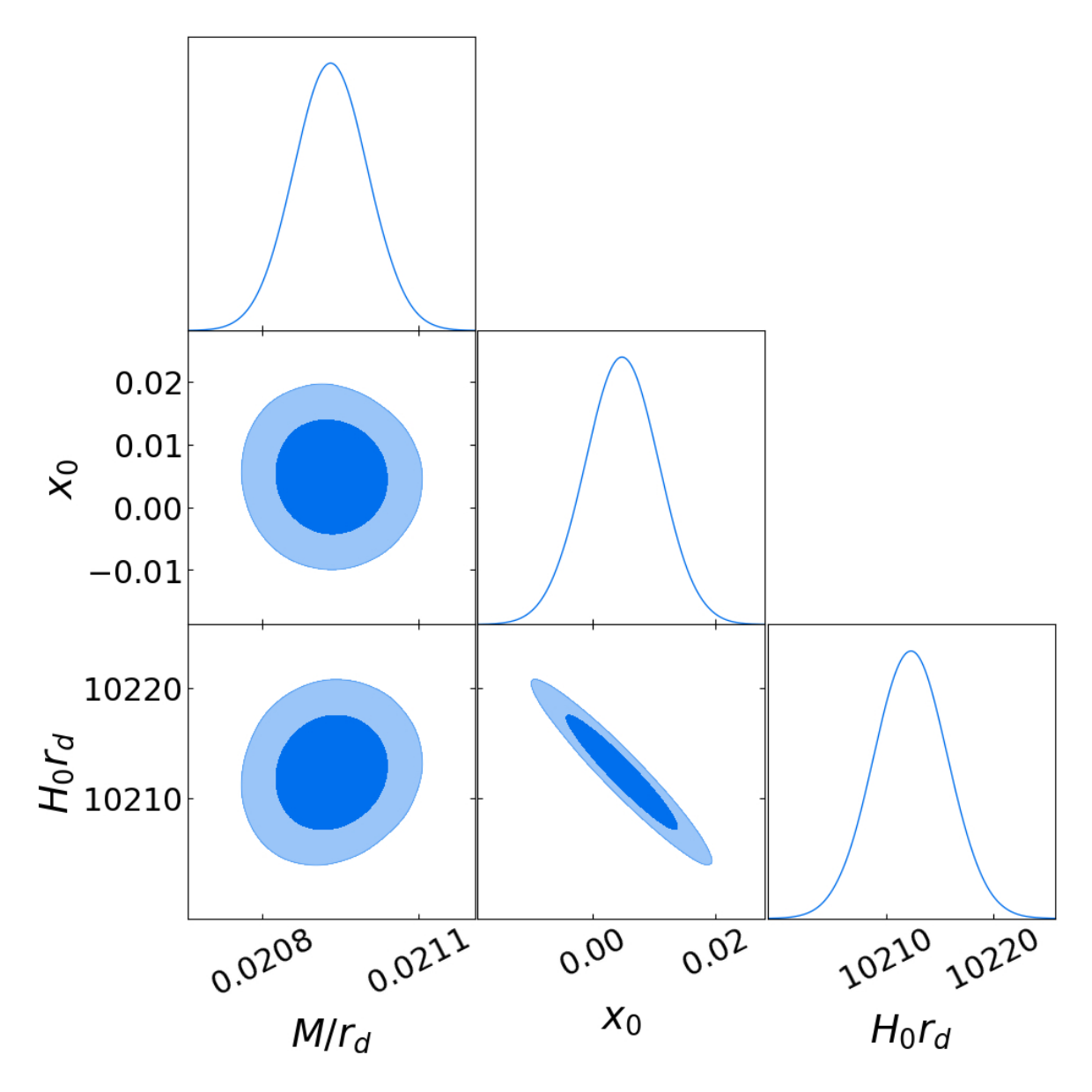}
\caption{NGC 6264}
\label{fig:6264}
\end{subfigure}
\begin{subfigure}{0.3\linewidth}
\includegraphics[width=\linewidth]{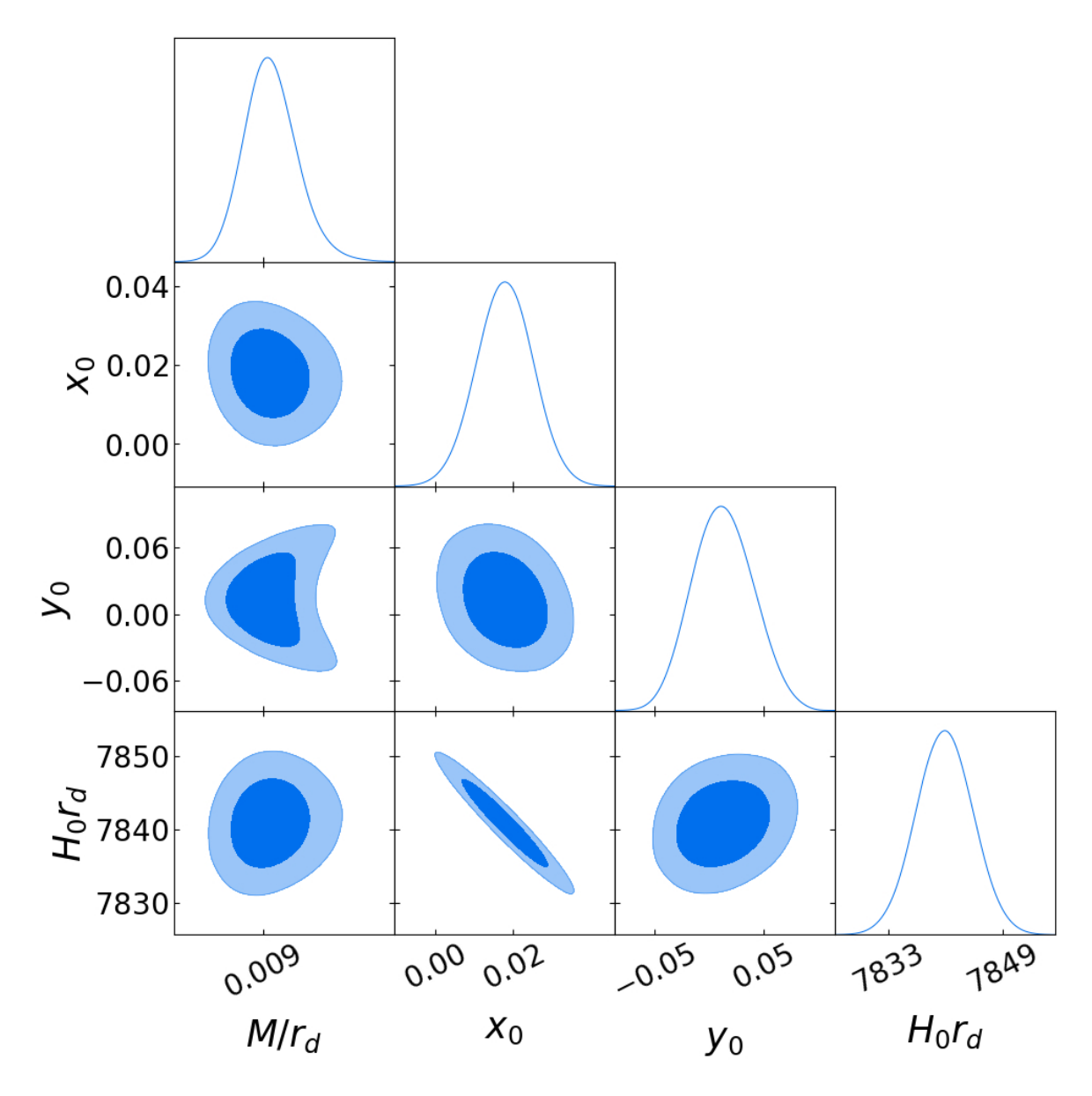}
\caption{NGC 6323}
\label{fig:6323}
\end{subfigure}
\begin{subfigure}{0.3\linewidth}
\includegraphics[width=\linewidth]{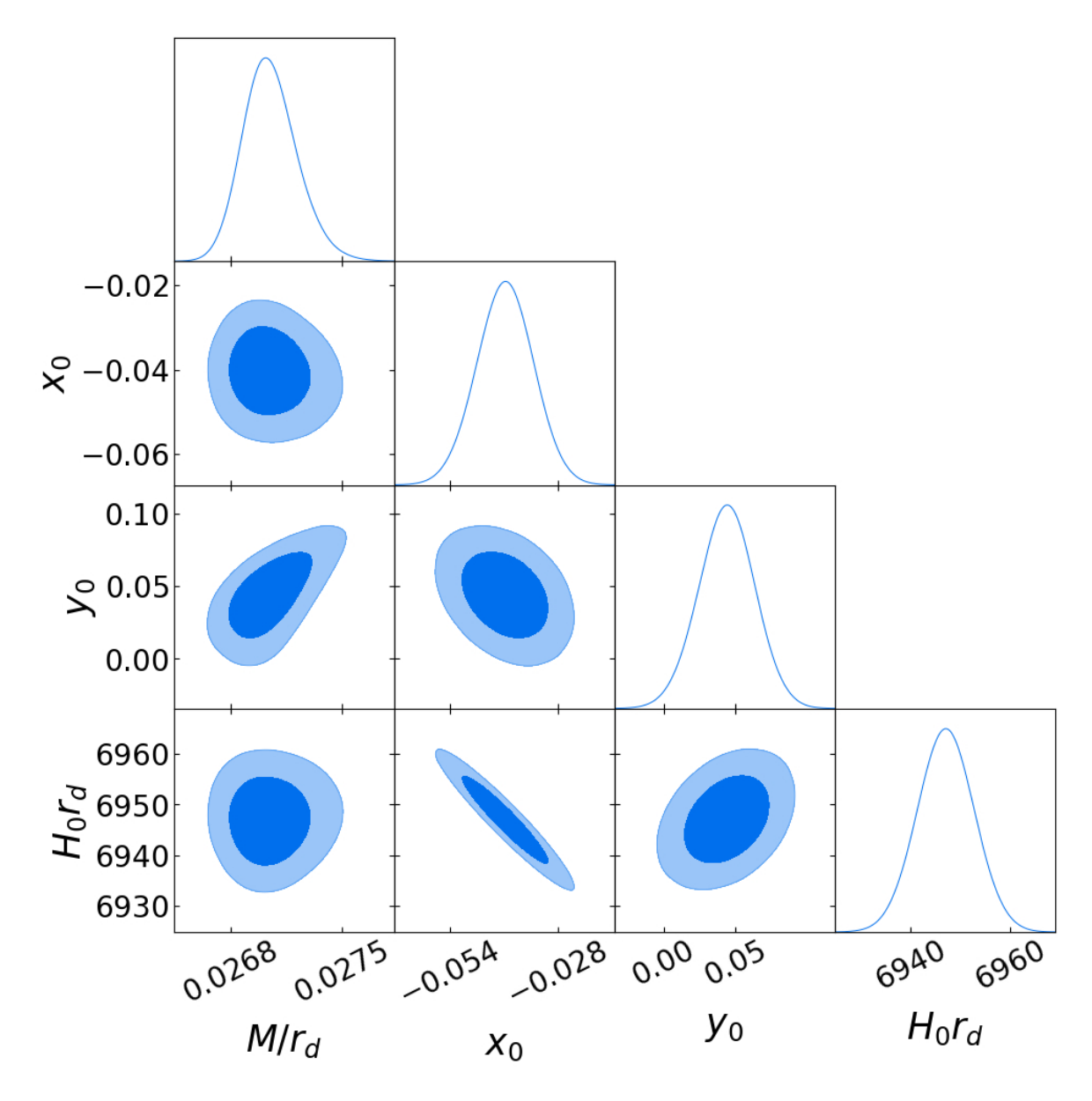}
\caption{CGCG 074-064}
\label{fig:cgcg}
\end{subfigure}
\caption{PDFs and confidence regions for the individual fits. The estimated parameters are the mass-to-distance ratio, the black hole angular position, and the product of the Hubble constant with the distance. The contour levels of the confidence regions correspond to 1$\sigma$
and 2$\sigma$.}
\label{fig:ind_pdf}
\end{figure*}

\begin{figure}
    \centering
    \includegraphics[width=0.9\linewidth]{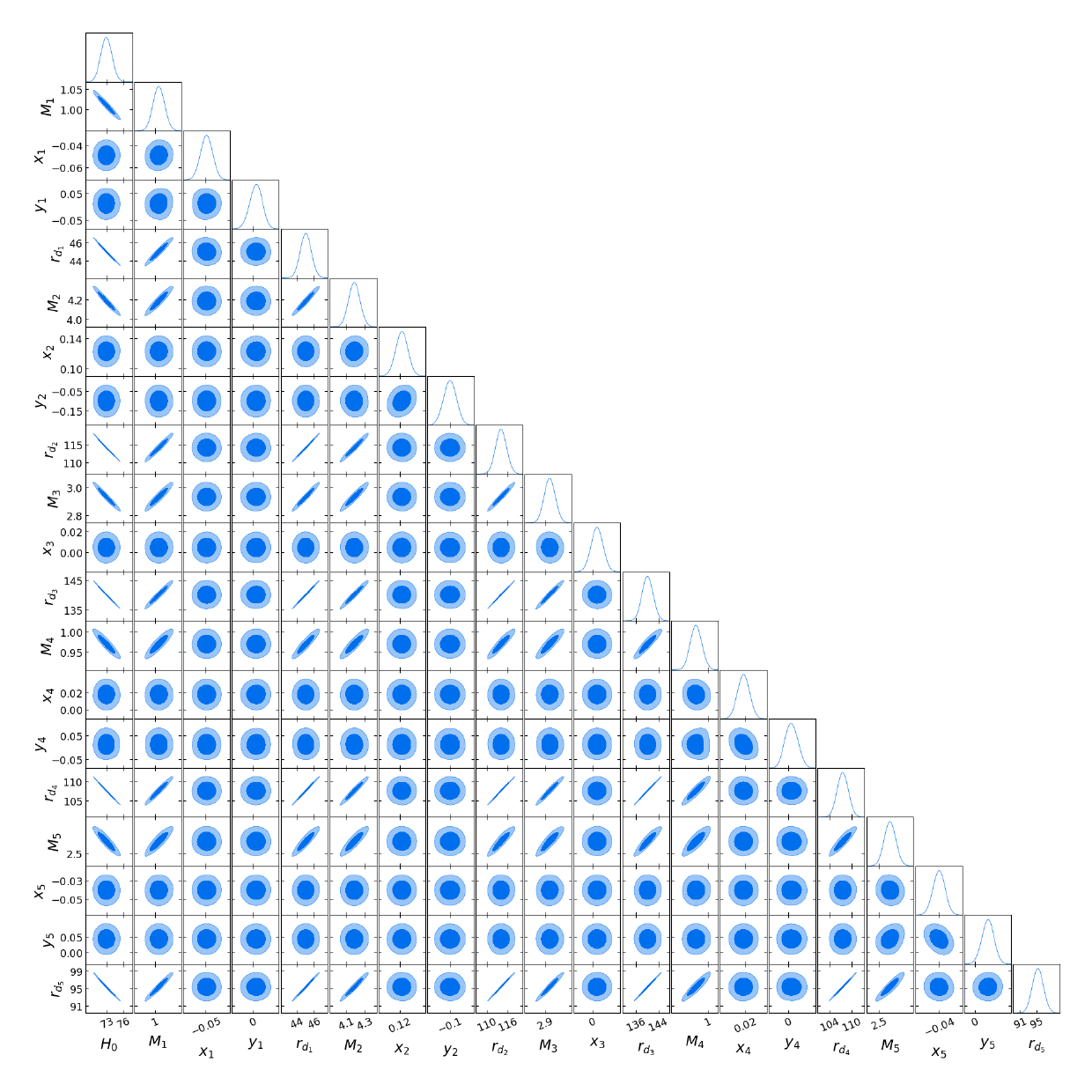}
    \caption{PDFs and confidence regions for the joint fit, where 20 parameters are estimated, corresponding the the mass, distance and position of each of the five AGN black holes, as well as the Hubble constant. }
    \label{fig:jointpdf}
\end{figure}

The resulting posterior probability distribution functions (PDFs) for the individual fits are shown in figure \ref{fig:ind_pdf}, and for the joint fit in figure \ref{fig:jointpdf}. In the joint fit, the strong correlation between the Hubble constant, the masses and distances from the five black holes becomes evident from the confidence regions. The galaxies then become correlated between each other because of the presence of the same Hubble constant in their redshift expressions, and the degeneracy between this constant with the black hole mass and distance.

\begin{table}[]
\begin{center}
\begin{tabular}{cccc}
\hline 
\multirow{2}{*}{Source} & \multirow{2}{*}{\begin{tabular}[c]{@{}c@{}}Distance to \\ closest maser\end{tabular}} & \multirow{2}{*}{$cZ_{g,SdS}$} & \multirow{2}{*}{$cZ_{kin\pm}$} \\
                        &                                                                                       &                        &                       \\
                        & (mas)                                                                                   & (km s$^{-1}$)                   & (km s$^{-1}$)                   \\
\hline \hline
\multirow{2}{*}{UGC 3789}   & \multirow{2}{*}{0.305}                                                                     & \multirow{2}{*}{3284.29}      & \multirow{2}{*}{-815.57}     \\
                        &                                                                                       &                        &                       \\
\hline
\multirow{2}{*}{NGC 5765}   & \multirow{2}{*}{0.538}                                                                     & \multirow{2}{*}{8317.73}      & \multirow{2}{*}{-799.42}     \\
                        &                                                                                       &                        &                       \\
\hline
\multirow{2}{*}{NGC 6264}   & \multirow{2}{*}{0.337}                                                                     & \multirow{2}{*}{10215.17}      & \multirow{2}{*}{-767.99}     \\
                        &                                                                                       &                        &                       \\
\hline
\multirow{2}{*}{NGC 6323}   & \multirow{2}{*}{0.215 }                                                                     & \multirow{2}{*}{7842.78}      & \multirow{2}{*}{625.33}     \\
                        &                                                                                       &                        &                       \\
\hline
\multirow{2}{*}{CGCG 074-064}   & \multirow{2}{*}{0.253}                                                                     & \multirow{2}{*}{6951.91}      & \multirow{2}{*}{-995.70}     \\
                        &                                                                                       &                        &        \\
\hline 
\end{tabular}
\caption{Gravitational and kinematic redshifts in SdS in terms of velocities for the closest maser to each black hole. The distance of the closest maser to the black hole is measured with respect to the estimated black hole position on the sky. The redshifts are reported in terms of velocities by using $v = cZ$.}
\label{tab:zgrav}
\end{center}
\end{table}

We compute the gravitational and the kinematic redshifts using equations \eqref{eq:zgrav} and \eqref{eq:zkin}, respectively, for each maser system. These redshifts are calculated for the innermost maser feature of each black hole, using the posterior values listed in table \ref{tab:ind_res}, and are reported in terms of velocity in table \ref{tab:zgrav}. The gravitational redshifts obtained here are in high contrast to those obtained from the Schwarzschild case, where the gravitational contribution is the order of $\sim 1$ km s$^{-1}$ \citep[see][]{Villaraos22}. This discrepancy arises because, in the SdS spacetime, the temporal component of the four-velocity, which is responsible for the gravitational redshift, explicitly depends on the cosmological constant and thus encodes the effect of the cosmic expansion. For the systems in UGC 3789, NGC 5765b, NGC 6264, and CGCG 074-064, the closest masers are blue features, hence the minus sign in their resulting kinematic redshift.

\section{Discussion and conclusions}
\label{sec:conclusion}

The approach presented in this work is based on the SdS metric, which incorporates the effect of a positive cosmological constant into a static, spherically symmetric geometry. We derived expressions for the redshift and blueshift of photons emitted by a particle at the midline and in circular motion around such black hole. By using the relation of the spacetime parameters in the frequency-shift expressions (mass, distance, black hole position, and the Hubble constant) with astrophysical observables (redshift and angular position of emission) we are able to estimate the former parameters corresponding to five megamaser AGN black holes, whose observational data are reported in the literature. 

More specifically, $M/r_d$ and $H_0r_d$ were estimated individually for each galaxy using flat priors, whereas $M$, $r_d$, and $H_0$ were estimated for the joint set of galaxies with a Gaussian prior on the Hubble constant only. The use of this Gaussian prior allows us to estimate the black hole masses, distances, and the Hubble constant separately; however, these parameters are not independent of each other, as they remain strongly correlated. The Gaussian prior on $H_0$ influences the posterior distributions of each mass and distance.

The results obtained from the individual and joint fits are consistent with previous studies \citep[e.g.][]{MCPXIII}. In the case of the individual fits, we obtain high precision in the estimated mass-to-distance ratio and the product of the Hubble constant with the distance, with uncertainties in the ranges between 0.3\% -- 0.7\% and  0.03\% -- 0.08\%, respectively. For the joint fit, the uncertainty in distance is 1.42\%, which remains notably small for extragalactic distance estimates. This uncertainty is largely influenced by the Gaussian prior on the Hubble constant, whose percent uncertainty is of similar magnitude. 

It is worth mentioning that the present framework, based on the SdS metric, serves as a first approximation, given that its validity is restricted to the late-time, $\Lambda$-dominated universe, and to distances well within the cosmological horizon. Hence, the inferred value of the Hubble constant should be considered as an effective cosmological parameter. This constant is introduced naturally through the relation \eqref{eq:hub_Lambda}, allowing the cosmological expansion to be encoded directly in the spacetime geometry rather than being introduced phenomenologically via the empirical Hubble law.  Moreover, the contribution of cosmic expansion is embedded in the gravitational redshift, which originates from the temporal component of the photon emitter's four-velocity, highlighting its intrinsically general relativistic nature. Thus, our results suggest a formalism based on general relativity that takes into account the contribution of cosmic expansion and differs from the standard Hubble law. 

In future work, the elimination of the Gaussian prior will be desirable in order to obtain parameter estimates that are independent of external measurements. This may be achieved by implementing the redshift rapidity \citep{MomenniaEPJ} into the model, allowing the quantities of black hole mass, its distance to Earth, and Hubble constant to decouple.

\appendix
\section{Redshift of a static particle}\label{app:Static}

In the simplest case of a static photon source in the Schwarzschild spacetime, the redshift detected by a distant  static observer reads
\begin{equation}
1 + Z_{Schw_{static}} = \sqrt{\frac{1}{1-2\tilde{M}}}.
\label{eq:staticSchw}
\end{equation}
This redshift arises purely from the gravitational potential generated by the black hole mass. To incorporate recessional motion due to the expansion of the universe, we add the redshift associated with the Hubble law as in \eqref{eq:z_schw_L}.

On the other hand, if we consider a static particle with respect to a SdS black hole, the redshift of its photons as detected by a distant observer with radial motion is
\begin{equation}
    1 + Z_{SdS_{static}} = \dfrac{1}{\sqrt{1 - 2\tilde{M} - \tilde{\Lambda} }} 
    \frac{1 - 2\bar{M} - \bar{\Lambda} }{ \sqrt{1-(9M^2\Lambda)^{1/3}}  - \sqrt{2\bar{M} + \bar{\Lambda} - (9M^2\Lambda)^{1/3} }  } .
\label{eq:staticSdS}
\end{equation}

\begin{figure}[]
        \centering
        \includegraphics[width=0.5\linewidth]{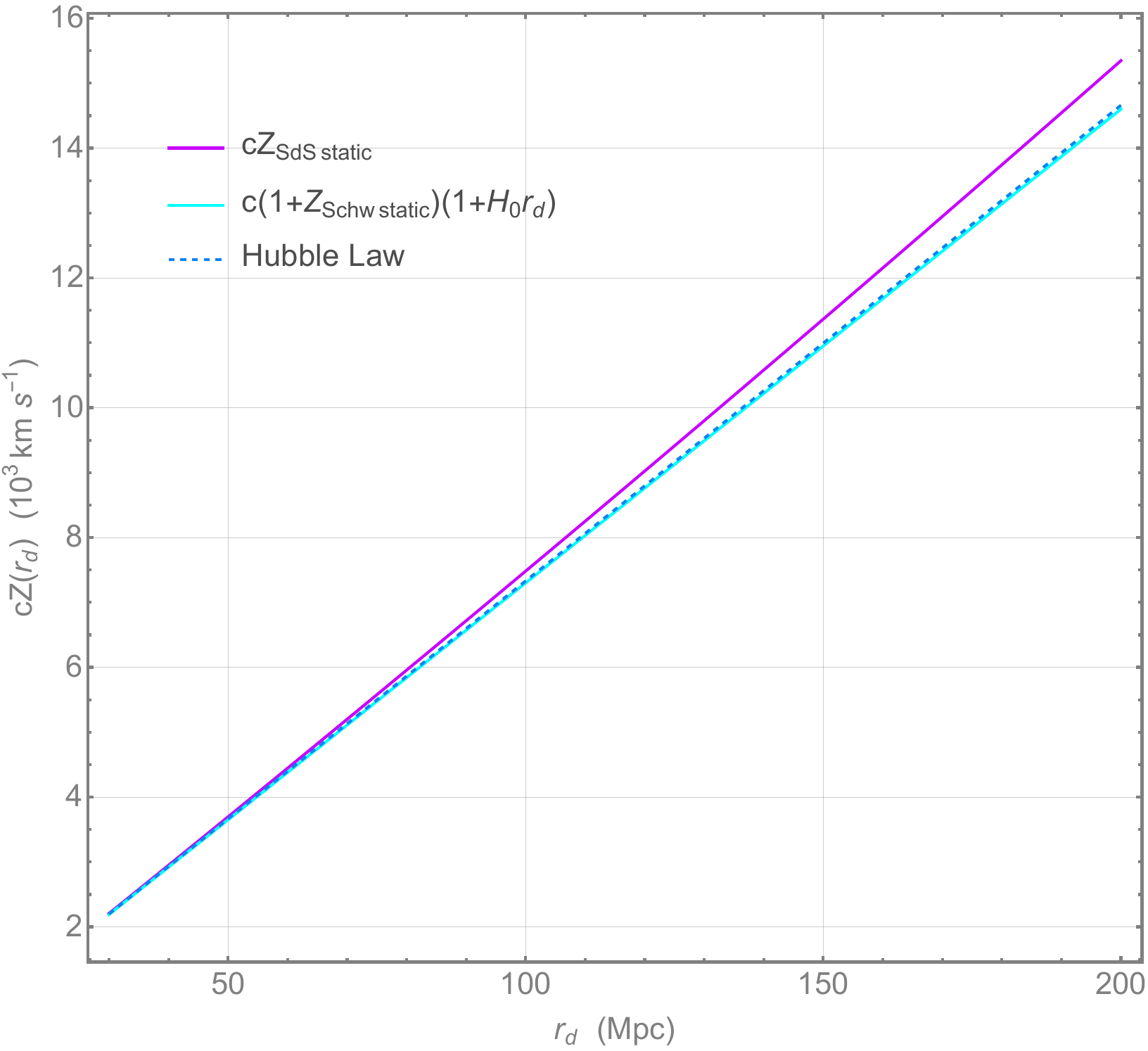}
        \caption{Plot of the redshift for a static particle in the SdS metric (purple), in the Schwarzschild metric with the Hubble law (blue), and the Hubble law (dotted line). }
        \label{fig:zstatic}
\end{figure}
In this case, the redshift comprises the gravitational effect due to the black hole presence as well as the cosmic expansion.
Notice that replacing circular motion with a static emitter eliminates the kinematic redshift and also modifies the gravitational term. Moreover, since there is no azimuthal motion, there is no blueshift in this case.
The behavior of this redshift is shown in figure \ref{fig:zstatic}, along with the redshift from the Schwarzschild metric composed with the Hubble law.

Interestingly, in the limit when  $\tilde{M}$, $\bar{M}$ and $\tilde{\Lambda}$ vanish, we recover the Hubble law (\ref{eq:zlambda}) as expected.

\acknowledgments

All authors are grateful to D. Martínez-Valera and A. González-Juarez for fruitful discussions and to L. Rojas Martínez and J. Tolentino Ramírez for technical support. The authors also thank the MCP researchers for making their data publicly available and appreciate support from FORDECYT-PRONACES-CONACYT under grant No. CF-MG-2558591. The authors thankfully acknowledge the computer resources, technical expertise and support provided by the Laboratorio Nacional de Supercómputo del Sureste de México, CONAHCYT member of the national network of laboratories. A.H.-A., M.M. and U.N. are grateful to SNII for support; A.H.-A. acknowledges a VIEP-BUAP grant, while D.V. acknowledge financial assistance from SECIHTI through the grant No. 1071008. M. M. was supported by SECIHTI through Estancias Posdoctorales por México Convocatoria 2023(1) under the postdoctoral Grant No. 1242413.


 \bibliographystyle{JHEP}
\bibliography{bib.bib}

@ARTICLE{Genzel87,
       author = {{Genzel}, R. and {Townes}, C.~H.},
        title = "{Physical conditions, dynamics, and mass distribution in the center of the galaxy.}",
      journal = {Ann. Rev. Astron. Astrophys.},
         year = 1987,
        month = jan,
       volume = {25},
        pages = {377-423},
          doi = {10.1146/annurev.aa.25.090187.002113},
       adsurl = {https://ui.adsabs.harvard.edu/abs/1987ARA&A..25..377G}
}

@ARTICLE{Ghez98,
       author = {{Ghez}, A.~M. and {Klein}, B.~L. and {Morris}, M. and {Becklin}, E.~E.},
        title = "{High Proper-Motion Stars in the Vicinity of Sagittarius A*: Evidence for a Supermassive Black Hole at the Center of Our Galaxy}",
      journal = {Astrophys. J. },
         year = 1998,
        month = dec,
       volume = {509},
       number = {2},
        pages = {678-686},
          doi = {10.1086/306528},
archivePrefix = {arXiv},
       eprint = {astro-ph/9807210},
 primaryClass = {astro-ph},
       adsurl = {https://ui.adsabs.harvard.edu/abs/1998ApJ...509..678G}
}

@ARTICLE{MomenniaEPJ,
       author = {{Momennia}, Mehrab and {Banerjee}, Pritam and {Herrera-Aguilar}, Alfredo and {Nucamendi}, Ulises},
        title = "{Schwarzschild black hole and redshift rapidity: a new approach towards measuring cosmic distances}",
      journal = {Eur. Phys. J. C},
         year = 2024,
        month = jun,
       volume = {84},
       number = {6},
          eid = {583},
        pages = {583},
          doi = {10.1140/epjc/s10052-024-12933-0},
archivePrefix = {arXiv},
       eprint = {2312.07426},
 primaryClass = {gr-qc},
       adsurl = {https://ui.adsabs.harvard.edu/abs/2024EPJC...84..583M}
}

@ARTICLE{Ghez08,
       author = {{Ghez}, A.~M. and {Salim}, S. and {Weinberg}, N.~N. and {Lu}, J.~R. and {Do}, T. and {Dunn}, J.~K. and {Matthews}, K. and {Morris}, M.~R. and {Yelda}, S. and {Becklin}, E.~E. and {Kremenek}, T. and {Milosavljevic}, M. and {Naiman}, J.},
        title = "{Measuring Distance and Properties of the Milky Way's Central Supermassive Black Hole with Stellar Orbits}",
      journal = {Astrophys. J. },
         year = 2008,
        month = dec,
       volume = {689},
       number = {2},
        pages = {1044-1062},
          doi = {10.1086/592738},
archivePrefix = {arXiv},
       eprint = {0808.2870},
 primaryClass = {astro-ph},
       adsurl = {https://ui.adsabs.harvard.edu/abs/2008ApJ...689.1044G}
}

@ARTICLE{Gillessen09,
       author = {{Gillessen}, S. and {Eisenhauer}, F. and {Trippe}, S. and {Alexander}, T. and {Genzel}, R. and {Martins}, F. and {Ott}, T.},
        title = "{Monitoring Stellar Orbits Around the Massive Black Hole in the Galactic Center}",
      journal = {Astrophys. J. },
         year = 2009,
        month = feb,
       volume = {692},
       number = {2},
        pages = {1075-1109},
          doi = {10.1088/0004-637X/692/2/1075},
archivePrefix = {arXiv},
       eprint = {0810.4674},
 primaryClass = {astro-ph},
       adsurl = {https://ui.adsabs.harvard.edu/abs/2009ApJ...692.1075G}
}

@ARTICLE{Lo2005,
       author = {{Lo}, K.~Y.},
        title = "{Mega-Masers and Galaxies}",
      journal = {Ann. Rev. Astron. Astrophys.},
         year = 2005,
        month = sep,
       volume = {43},
       number = {1},
        pages = {625-676},
          doi = {10.1146/annurev.astro.41.011802.094927},
       adsurl = {https://ui.adsabs.harvard.edu/abs/2005ARA&A..43..625L}
}

@ARTICLE{Herrnstein99,
       author = {{Herrnstein}, J.~R. and {Moran}, J.~M. and {Greenhill}, L.~J. and {Diamond}, P.~J. and {Inoue}, M. and {Nakai}, N. and {Miyoshi}, M. and {Henkel}, C. and {Riess}, A.},
        title = "{A geometric distance to the galaxy NGC4258 from orbital motions in a nuclear gas disk}",
      journal = {Nature},
         year = 1999,
        month = aug,
       volume = {400},
       number = {6744},
        pages = {539-541},
          doi = {10.1038/22972},
archivePrefix = {arXiv},
       eprint = {astro-ph/9907013},
 primaryClass = {astro-ph},
       adsurl = {https://ui.adsabs.harvard.edu/abs/1999Natur.400..539H}
}

@ARTICLE{MCPIV,
       author = {{Reid}, M.~J. and {Braatz}, J.~A. and {Condon}, J.~J. and {Lo}, K.~Y. and {Kuo}, C.~Y. and {Impellizzeri}, C.~M.~V. and {Henkel}, C.},
        title = "{The Megamaser Cosmology Project. IV. A Direct Measurement of the Hubble Constant from UGC 3789}",
      journal = {Astrophys. J.},
         year = 2013,
        month = apr,
       volume = {767},
       number = {2},
          eid = {154},
        pages = {154},
          doi = {10.1088/0004-637X/767/2/154},
archivePrefix = {arXiv},
       eprint = {1207.7292},
 primaryClass = {astro-ph.CO},
       adsurl = {https://ui.adsabs.harvard.edu/abs/2013ApJ...767..154R}
}

@article{Herrera2015,
    author = "Herrera-Aguilar, Alfredo and Nucamendi, Ulises",
    title = "{Kerr black hole parameters in terms of the redshift/blueshift of photons emitted by geodesic particles}",
    eprint = "1506.05182",
    archivePrefix = "arXiv",
    primaryClass = "gr-qc",
    doi = "10.1103/PhysRevD.92.045024",
    journal = "Phys. Rev. D",
    volume = "92",
    number = "4",
    pages = "045024",
    year = "2015"
}

@article{Momennia2023,
  title = {Kerr black hole in de Sitter spacetime and observational redshift: Toward a new method to measure the Hubble constant},
  author = {Momennia, Mehrab and Herrera-Aguilar, Alfredo and Nucamendi, Ulises},
  journal = {Phys. Rev. D},
  volume = {107},
  issue = {10},
  pages = {104041},
  numpages = {16},
  year = {2023},
  month = {May},
  publisher = {American Physical Society},
  doi = {10.1103/PhysRevD.107.104041},
  url = {https://link.aps.org/doi/10.1103/PhysRevD.107.104041}
}

@article{Gerardo,
    author = {Morales-Herrera, Gerardo and Ortega-Ruiz, Pablo and Momennia, Mehrab and Herrera-Aguilar, Alfredo},
    title = {{Mass, charge, and distance to Reissner\textendash{}Nordstr\"om black hole in terms of directly measurable quantities}},
    eprint = "2401.07112",
    archivePrefix = "arXiv",
    primaryClass = "gr-qc",
    doi = "10.1140/epjc/s10052-024-12880-w",
    journal = "Eur. Phys. J. C",
    volume = "84",
    number = "5",
    pages = "525",
    year = "2024"
}

@article{Diego2024,
    author ={Mart{\'\i}nez-Valera, Diego A. and Momennia, Mehrab and Herrera-Aguilar, Alfredo},
    title = "{Observational redshift from general spherically symmetric black holes}",
    eprint = "2311.17993",
    archivePrefix = "arXiv",
    primaryClass = "gr-qc",
    doi = "10.1140/epjc/s10052-025-14114-z",
    journal = "Eur. Phys. J. C",
    volume = "84",
    number = "3",
    pages = "288",
    year = "2024"
}

@ARTICLE{1918Kottler,
       author = {{Kottler}, F.},
        title = "{{\"U}ber die physikalischen Grundlagen der Einsteinschen Gravitationstheorie}",
      journal = {Annalen der Physik},
         year = 1918,
        month = jan,
       volume = {361},
       number = {14},
        pages = {401-462},
          doi = {10.1002/andp.19183611402},
       adsurl = {https://ui.adsabs.harvard.edu/abs/1918AnP...361..401K}
}

@article{Banerjee2022,
  author      = {{Banerjee}, Pritam and {Herrera-Aguilar}, Alfredo  and {Momennia}, Mehrab and {Nucamendi}, Ulises
                },
  title       = {{Mass and Spin of Kerr Black Holes in Terms of Observational Quantities: The Dragging Effect on the Redshift}},
  journal     = {Phys. Rev. D},
  volume      = {105},
  pages       = {124037},
  month       = {Jun},
  year        = {2022},
  doi         = {10.1103/PhysRevD.105.124037},
}

@ARTICLE{Claussen84,
       author = {{Claussen}, M.~J. and {Heiligman}, G.~M. and {Lo}, K.~Y.},
        title = "{Water-vapour maser emission from galactic nuclei}",
      journal = {Nature},
         year = 1984,
        month = jul,
       volume = {310},
       number = {5975},
        pages = {298-300},
          doi = {10.1038/310298a0},
       adsurl = {https://ui.adsabs.harvard.edu/abs/1984Natur.310..298C}
}

@article{MCPIX,
author = {Gao, F. and Braatz, J. and Reid, M. and Condon, James and Greene, J. and Henkel, C. and Impellizzeri, C. and Lo, K. and Kuo, Chien-Yin and Pesce, D. and Wagner, J. and Zhao, W.},
year = {2017},
month = {10},
pages = {52},
title = {The Megamaser Cosmology Project. IX. Black hole masses for three maser galaxies},
volume = {834},
journal = {Astrophys. J.},
doi = {10.3847/1538-4357/834/1/52}
}

@article{MCPXII,
author = {Kuo, C.~Y. and Braatz, J and Impellizzeri, C. and Gao, F and Pesce, D and Reid, M and Condon, James and Kamali, Fateme and Henkel, C and Greene, J},
year = {2020},
month = {09},
pages = {1609-1627},
title = {The Megamaser Cosmology Project - XII. VLBI imaging of H2O maser emission in three active galaxies and the effect of AGN winds on disc dynamics},
volume = {498},
journal = {Mon. Not. R. Astron. Soc.},
doi = {10.1093/mnras/staa2260}
}

@article{MCPVIII,
author = {Gao, F. and Braatz, J. and Reid, M. and Lo, K. and Condon, James and Henkel, C. and Kuo, Chien-Yin and Impellizzeri, C. and Pesce, D. and Zhao, W.},
year = {2016},
month = {11},
pages = {128},
title = {The Megamaser Cosmology Project. VIII. A Geometric Distance to NGC 5765b},
volume = {817},
journal = {Astrophys. J.},
doi = {10.3847/0004-637X/817/2/128}
}

@article{MCPXI,
author = {Pesce, D. and Braatz, J. and Reid, M. and Condon, James and Gao, F. and Henkel, C. and Kuo, Chien-Yin and Lo, K. and Zhao, W.},
year = {2020},
month = {01},
pages = {},
title = {The Megamaser Cosmology Project. XI. A geometric distance to CGCG 074-064},
volume = {890},
journal = {Astrophys. J.},
doi = {10.3847/1538-4357/ab6bcd}
}

@ARTICLE{MCPXIII,
       author = {{Pesce}, D. and {Braatz}, J.~A. and {Reid}, M.~J. and {Riess}, A.~G. and {Scolnic}, D. and {Condon}, J.~J. and {Gao}, F. and {Henkel}, C. and {Impellizzeri}, C.~M.~V. and {Kuo}, C.~Y. and {Lo}, K.~Y.},
        title = "{The Megamaser Cosmology Project. XIII. Combined Hubble Constant Constraints}",
      journal = {Astrophys. J. Lett.},
         year = 2020,
        month = mar,
       volume = {891},
       number = {1},
          eid = {L1},
        pages = {L1},
          doi = {10.3847/2041-8213/ab75f0},
archivePrefix = {arXiv},
       eprint = {2001.09213},
 primaryClass = {astro-ph.CO},
       adsurl = {https://ui.adsabs.harvard.edu/abs/2020ApJ...891L...1P}
}

@ARTICLE{Riess22,
       author = {{Riess}, Adam G. and {Yuan}, Wenlong and {Macri}, Lucas M. and {Scolnic}, Dan and {Brout}, Dillon and {Casertano}, Stefano and {Jones}, David O. and {Murakami}, Yukei and {Anand}, Gagandeep S. and {Breuval}, Louise and {Brink}, Thomas G. and {Filippenko}, Alexei V. and {Hoffmann}, Samantha and {Jha}, Saurabh W. and {D'arcy Kenworthy}, W. and {Mackenty}, John and {Stahl}, Benjamin E. and {Zheng}, WeiKang},
        title = "{A Comprehensive Measurement of the Local Value of the Hubble Constant with 1 km s$^{-1}$ Mpc$^{-1}$ Uncertainty from the Hubble Space Telescope and the SH0ES Team}",
      journal = {Astrophys. J. Lett.},
         year = 2022,
        month = jul,
       volume = {934},
       number = {1},
          eid = {L7},
        pages = {L7},
          doi = {10.3847/2041-8213/ac5c5b},
archivePrefix = {arXiv},
       eprint = {2112.04510},
 primaryClass = {astro-ph.CO},
       adsurl = {https://ui.adsabs.harvard.edu/abs/2022ApJ...934L...7R}
}

@ARTICLE{Riess24,
       author = {{Riess}, Adam G. and {Scolnic}, Dan and {Anand}, Gagandeep S. and {Breuval}, Louise and {Casertano}, Stefano and {Macri}, Lucas M. and {Li}, Siyang and {Yuan}, Wenlong and {Huang}, Caroline D. and {Jha}, Saurabh and {Murakami}, Yukei S. and {Beaton}, Rachael and {Brout}, Dillon and {Wu}, Tianrui and {Addison}, Graeme E. and {Bennett}, Charles and {Anderson}, Richard I. and {Filippenko}, Alexei V. and {Carr}, Anthony},
        title = "{JWST Validates HST Distance Measurements: Selection of Supernova Subsample Explains Differences in JWST Estimates of Local H $_{0}$}",
      journal = {Astrophys. J. },
         year = 2024,
        month = dec,
       volume = {977},
       number = {1},
          eid = {120},
        pages = {120},
          doi = {10.3847/1538-4357/ad8c21},
archivePrefix = {arXiv},
       eprint = {2408.11770},
 primaryClass = {astro-ph.CO},
       adsurl = {https://ui.adsabs.harvard.edu/abs/2024ApJ...977..120R}
}

@article{Nucamendi21,
    author = "Nucamendi, Ulises and Herrera-Aguilar, Alfredo and Lizardo-Castro, Raul and Cruz, Omar Lopez",
    title = "{Toward the Gravitational Redshift Detection in NGC 4258 and the Estimation of Its Black Hole Mass-to-distance Ratio}",
    eprint = "2012.05487",
    archivePrefix = "arXiv",
    primaryClass = "gr-qc",
    doi = "10.3847/2041-8213/ac151b",
    journal = "Astrophys. J. Lett.",
    volume = "917",
    number = "1",
    pages = "L14",
    year = "2021"
}

@article{Dainotti25,
    author = "Dainotti, M. G. and others",
    title = "{A New Master Supernovae Ia sample and the investigation of the Hubble tension}",
    eprint = "2501.11772",
    archivePrefix = "arXiv",
    primaryClass = "astro-ph.CO",
    reportNumber = "KEK-TH-2711, KEK-Cosmo-0378",
    doi = "10.1016/j.jheap.2025.100405",
    journal = "JHEAp",
    volume = "48",
    pages = "100405",
    year = "2025"
}

@ARTICLE{Benisty24,
       author = {{Benisty}, David and {Vasak}, David and {Struckmeier}, J{\"u}rgen and {Stoecker}, Horst},
        title = "{Bounding the cosmological constant using galactic rotation curves from SPARC dataset}",
      journal = {Phys. Rev. D},
         year = 2024,
        month = sep,
       volume = {110},
       number = {6},
          eid = {063028},
        pages = {063028},
          doi = {10.1103/PhysRevD.110.063028},
archivePrefix = {arXiv},
       eprint = {2405.16650},
 primaryClass = {astro-ph.CO},
       adsurl = {https://ui.adsabs.harvard.edu/abs/2024PhRvD.110f3028B}
}

@ARTICLE{Villaraos22,
       author = {{Villaraos}, D. and {Herrera-Aguilar}, A. and {Nucamendi}, U. and {Gonz{\'a}lez-Ju{\'a}rez}, G. and {Lizardo-Castro}, R.},
        title = "{A general relativistic mass-to-distance ratio for a set of megamaser AGN black holes}",
      journal = {Mon. Not. R. Astron. Soc.},
         year = 2022,
        month = dec,
       volume = {517},
       number = {3},
        pages = {4213-4219},
          doi = {10.1093/mnras/stac2973},
archivePrefix = {arXiv},
       eprint = {2207.06594},
 primaryClass = {astro-ph.GA},
       adsurl = {https://ui.adsabs.harvard.edu/abs/2022MNRAS.517.4213V}
}

@article{Adri24,
    author = "Gonz\'alez-Ju\'arez, A. and Momennia, M. and Villalobos-Ram\'\i{}rez, A. and Herrera-Aguilar, A.",
    title = "{Estimating the mass-to-distance ratio for a set of megamaser AGN black holes by employing a general relativistic method}",
    eprint = "2211.06486",
    archivePrefix = "arXiv",
    primaryClass = "astro-ph.GA",
    doi = "10.1051/0004-6361/202450098",
    journal = "Astron. Astrophys.",
    volume = "689",
    pages = "A205",
    year = "2024"
}

@article{diego25,
   author={Diego A. Martínez-Valera and Alfredo Herrera-Aguilar},
    title={Parameter estimation of nonsingular black holes in conformal gravity using megamaser observational data from NGC 4258}, 
    eprint={2504.04588},
    archivePrefix={arXiv},
    primaryClass={gr-qc},
    doi = {10.1140/epjc/s10052-025-15208-4},
    journal = "Eur. Phys. J. C",
    volume = "85",
    pages = "1472",
    year = "2025"
}

@article{Iorio25,
  title = {Lense-Thirring effect at work in $\mathrm{M}8{7}^{*}$},
  author = {Iorio, Lorenzo},
  journal = {Phys. Rev. D},
  volume = {111},
  issue = {4},
  pages = {044035},
  numpages = {12},
  year = {2025},
  month = {Feb},
  publisher = {American Physical Society},
  doi = {10.1103/PhysRevD.111.044035},
  url = {https://link.aps.org/doi/10.1103/PhysRevD.111.044035}
}

@article{Adri25,
author = {González-Juárez, Adriana and Herrera-Aguilar, Alfredo},
year = {2025},
title = {Reviewing the GR Method for Estimating Black Hole Parameters of Megamaser Systems},
journal = {Astron. Nachr.},
volume = {346},
number = {3-4},
pages = {e20250016},
doi = {https://doi.org/10.1002/asna.20250016},
url = {https://onlinelibrary.wiley.com/doi/abs/10.1002/asna.20250016},
eprint = {https://onlinelibrary.wiley.com/doi/pdf/10.1002/asna.20250016}
}

@ARTICLE{mcpii,
       author = {{Braatz}, J.~A. and {Reid}, M.~J. and {Humphreys}, E.~M.~L. and {Henkel}, C. and {Condon}, J.~J. and {Lo}, K.~Y.},
        title = "{The Megamaser Cosmology Project. II. The Angular-diameter Distance to UGC 3789}",
      journal = {Astrophys. J. },
         year = 2010,
        month = aug,
       volume = {718},
       number = {2},
        pages = {657-665},
          doi = {10.1088/0004-637X/718/2/657},
archivePrefix = {arXiv},
       eprint = {1005.1955},
 primaryClass = {astro-ph.CO},
       adsurl = {https://ui.adsabs.harvard.edu/abs/2010ApJ...718..657B}
}

@article{mcpvii,
    author = "Pesce, D. W. and Braatz, J. A. and Condon, J. J. and Gao, F. and Henkel, C. and Litzinger, E. and Lo, K. Y. and Reid, M. J.",
    title = "{The Megamaser Cosmology Project. VII. Investigating disk physics using spectral monitoring observations}",
    eprint = "1507.07904",
    archivePrefix = "arXiv",
    primaryClass = "astro-ph.GA",
    doi = "10.1088/0004-637X/810/1/65",
    journal = "Astrophys. J.",
    volume = "810",
    number = "1",
    pages = "65",
    year = "2015"
}

@ARTICLE{mcpv,
       author = {{Kuo}, C.~Y. and {Braatz}, J.~A. and {Reid}, M.~J. and {Lo}, K.~Y. and {Condon}, J.~J. and {Impellizzeri}, C.~M.~V. and {Henkel}, C.},
        title = "{The Megamaser Cosmology Project. V. An Angular-diameter Distance to NGC 6264 at 140 Mpc}",
      journal = {Astrophys. J. },
         year = 2013,
        month = apr,
       volume = {767},
       number = {2},
          eid = {155},
        pages = {155},
          doi = {10.1088/0004-637X/767/2/155},
archivePrefix = {arXiv},
       eprint = {1207.7273},
 primaryClass = {astro-ph.CO},
       adsurl = {https://ui.adsabs.harvard.edu/abs/2013ApJ...767..155K}
}

@ARTICLE{mcpvi,
       author = {{Kuo}, C.~Y. and {Braatz}, J.~A. and {Lo}, K.~Y. and {Reid}, M.~J. and {Suyu}, S.~H. and {Pesce}, D.~W. and {Condon}, J.~J. and {Henkel}, C. and {Impellizzeri}, C.~M.~V.},
        title = "{The Megamaser Cosmology Project. VI. Observations of NGC 6323}",
      journal = {Astrophys. J. },
         year = 2015,
        month = feb,
       volume = {800},
       number = {1},
          eid = {26},
        pages = {26},
          doi = {10.1088/0004-637X/800/1/26},
archivePrefix = {arXiv},
       eprint = {1411.5106},
 primaryClass = {astro-ph.GA},
       adsurl = {https://ui.adsabs.harvard.edu/abs/2015ApJ...800...26K}
}

@article{Sharif2016,
    author = "Sharif, M. and Iftikhar, Sehrish",
    title = "{Dynamics of Particles Near Black Hole with Higher Dimensions}",
    eprint = "1607.03507",
    archivePrefix = "arXiv",
    primaryClass = "gr-qc",
    doi = "10.1140/epjc/s10052-016-4244-0",
    journal = "Eur. Phys. J. C",
    volume = "76",
    number = "7",
    pages = "404",
    year = "2016"
}

@article{Shankar2018,
    author = "Shankar Kuniyal, Ravi and Uniyal, Rashmi and Biswas, Anindya and Nandan, Hemwati and Purohit, K. D.",
    title = "{Null geodesics and red\textendash{}blue shifts of photons emitted from geodesic particles around a noncommutative black hole space\textendash{}time}",
    eprint = "1703.07921",
    archivePrefix = "arXiv",
    primaryClass = "gr-qc",
    doi = "10.1142/S0217751X18500987",
    journal = "Int. J. Mod. Phys. A",
    volume = "33",
    number = "16",
    pages = "1850098",
    year = "2018"
}

@article{Uniyal17,
    author = "Uniyal, Rashmi and Nandan, Hemwati and Purohit, K. D.",
    title = "{Null geodesics and observables around the Kerr\textendash{}Sen black hole}",
    eprint = "1703.07510",
    archivePrefix = "arXiv",
    primaryClass = "gr-qc",
    doi = "10.1088/1361-6382/aa9ad9",
    journal = "Class. Quant. Grav.",
    volume = "35",
    number = "2",
    pages = "025003",
    year = "2018"
}

@article{Sheoran17,
    author = "Sheoran, Pankaj and Herrera-Aguilar, Alfredo and Nucamendi, Ulises",
    title = "{Mass and spin of a Kerr black hole in modified gravity and a test of the Kerr black hole hypothesis}",
    eprint = "1712.03344",
    archivePrefix = "arXiv",
    primaryClass = "gr-qc",
    doi = "10.1103/PhysRevD.97.124049",
    journal = "Phys. Rev. D",
    volume = "97",
    number = "12",
    pages = "124049",
    year = "2018"
}

@article{Kraniotis19,
    author = "Kraniotis, G. V.",
    title = "{Gravitational redshift/blueshift of light emitted by geodesic test particles, frame-dragging and pericentre-shift effects, in the Kerr\textendash{}Newman\textendash{}de Sitter and Kerr\textendash{}Newman black hole geometries}",
    eprint = "1912.10320",
    archivePrefix = "arXiv",
    primaryClass = "gr-qc",
    doi = "10.1140/epjc/s10052-021-08911-5",
    journal = "Eur. Phys. J. C",
    volume = "81",
    number = "2",
    pages = "147",
    year = "2021"
}

@article{Lopez21,
    author = "L\'opez, L. A. and Bret\'on, Nora",
    title = "{Redshift of light emitted by particles orbiting a black hole immersed in a strong magnetic field}",
    eprint = "2104.00840",
    archivePrefix = "arXiv",
    primaryClass = "gr-qc",
    doi = "10.1007/s10509-021-03961-3",
    journal = "Astrophys. Space Sci.",
    volume = "366",
    number = "6",
    pages = "55",
    year = "2021"
}

@article{Debnath21,
    author = "Debnath, Ujjal",
    title = "{Motion and collision of particles near Plebanski-Demianski black hole: Shadow and gravitational lensing}",
    doi = "10.1016/j.cjph.2020.09.037",
    journal = "Chin. J. Phys.",
    volume = "70",
    pages = "213--231",
    year = "2021"
}

@article{Giambo22,
    author = "Giamb\`o, Roberto and Luongo, Orlando and Mauro, Lorenza",
    title = "{Red and blue shift in spherical and axisymmetric spacetimes and astrophysical constraints}",
    eprint = "2206.14043",
    archivePrefix = "arXiv",
    primaryClass = "gr-qc",
    doi = "10.1140/epjp/s13360-022-02803-7",
    journal = "Eur. Phys. J. Plus",
    volume = "137",
    number = "5",
    pages = "612",
    year = "2022"
}

@article{Fu22,
    author = "Fu, Qi-Ming and Zhang, Xin",
    title = "{Probing a polymerized black hole with the frequency shifts of photons}",
    eprint = "2212.11474",
    archivePrefix = "arXiv",
    primaryClass = "gr-qc",
    doi = "10.1103/PhysRevD.107.064019",
    journal = "Phys. Rev. D",
    volume = "107",
    number = "6",
    pages = "064019",
    year = "2023"
}

@article{Mustafa22,
    author = "Mustafa, G. and Hussain, Ibrar and Liu, Wu-Ming",
    title = "{Quasi-periodic oscillations of test particles and red\textendash{}blue shifts of photons in the charged-Kiselev black hole with cloud of strings}",
    doi = "10.1016/j.cjph.2022.04.023",
    journal = "Chin. J. Phys.",
    volume = "80",
    pages = "148--166",
    year = "2022"
}

@article{Freedman24,
    author = {Freedman, Wendy L. and Madore, Barry F. and Hoyt, Taylor J. and Jang, In Sung and Lee, Abigail J. and Owens, Kayla A.},
    title = {{Status Report on the Chicago-Carnegie Hubble Program (CCHP): Measurement of the Hubble Constant Using the Hubble and James Webb Space Telescopes}},
    eprint = {2408.06153},
    archivePrefix = {arXiv},
    primaryClass = {astro-ph.CO},
    doi = {10.3847/1538-4357/adce78},
    journal = {Astrophys. J. },
    volume = {985},
    number = {2},
    pages = {203},
    year = {2025}
}

@article{Lee24,
    author = {Lee, Abigail J. and Freedman, Wendy L. and Madore, Barry F. and Jang, In Sung and Owens, Kayla A. and Hoyt, Taylor J.},
    title = {{The Chicago\textendash{}Carnegie Hubble Program: The JWST J-region Asymptotic Giant Branch Extragalactic Distance Scale*}},
    eprint = {2408.03474},
    archivePrefix = {arXiv},
    primaryClass = {astro-ph.GA},
    doi = {10.3847/1538-4357/adc8a1},
    journal = {Astrophys. J. },
    volume = {985},
    number = {2},
    pages = {182},
    year = {2025}
}

@ARTICLE{emcee,
       author = {{Foreman-Mackey}, Daniel and {Hogg}, David W. and {Lang}, Dustin and {Goodman}, Jonathan},
        title = "{emcee: The MCMC Hammer}",
      journal = {Publ. Astron. Soc. Pac.},
         year = 2013,
        month = mar,
       volume = {125},
       number = {925},
        pages = {306},
          doi = {10.1086/670067},
archivePrefix = {arXiv},
       eprint = {1202.3665},
 primaryClass = {astro-ph.IM},
       adsurl = {https://ui.adsabs.harvard.edu/abs/2013PASP..125..306F}
}

@ARTICLE{Planck20,
       author = {{Planck Collaboration} and {Aghanim}, N. and {Akrami}, Y. and {Ashdown}, M. and {Aumont}, J. and {Baccigalupi}, C. and {Ballardini}, M. and {Banday}, A.~J. and {Barreiro}, R.~B. and {Bartolo}, N. and {Basak}, S. and {Battye}, R. and {Benabed}, K. and {Bernard}, J. -P. and {Bersanelli}, M. and {Bielewicz}, P. and {Bock}, J.~J. and {Bond}, J.~R. and {Borrill}, J. and {Bouchet}, F.~R. and {Boulanger}, F. and {Bucher}, M. and {Burigana}, C. and {Butler}, R.~C. and {Calabrese}, E. and {Cardoso}, J. -F. and {Carron}, J. and {Challinor}, A. and {Chiang}, H.~C. and {Chluba}, J. and {Colombo}, L.~P.~L. and {Combet}, C. and {Contreras}, D. and {Crill}, B.~P. and {Cuttaia}, F. and {de Bernardis}, P. and {de Zotti}, G. and {Delabrouille}, J. and {Delouis}, J. -M. and {Di Valentino}, E. and {Diego}, J.~M. and {Dor{\'e}}, O. and {Douspis}, M. and {Ducout}, A. and {Dupac}, X. and {Dusini}, S. and {Efstathiou}, G. and {Elsner}, F. and {En{\ss}lin}, T.~A. and {Eriksen}, H.~K. and {Fantaye}, Y. and {Farhang}, M. and {Fergusson}, J. and {Fernandez-Cobos}, R. and {Finelli}, F. and {Forastieri}, F. and {Frailis}, M. and {Fraisse}, A.~A. and {Franceschi}, E. and {Frolov}, A. and {Galeotta}, S. and {Galli}, S. and {Ganga}, K. and {G{\'e}nova-Santos}, R.~T. and {Gerbino}, M. and {Ghosh}, T. and {Gonz{\'a}lez-Nuevo}, J. and {G{\'o}rski}, K.~M. and {Gratton}, S. and {Gruppuso}, A. and {Gudmundsson}, J.~E. and {Hamann}, J. and {Handley}, W. and {Hansen}, F.~K. and {Herranz}, D. and {Hildebrandt}, S.~R. and {Hivon}, E. and {Huang}, Z. and {Jaffe}, A.~H. and {Jones}, W.~C. and {Karakci}, A. and {Keih{\"a}nen}, E. and {Keskitalo}, R. and {Kiiveri}, K. and {Kim}, J. and {Kisner}, T.~S. and {Knox}, L. and {Krachmalnicoff}, N. and {Kunz}, M. and {Kurki-Suonio}, H. and {Lagache}, G. and {Lamarre}, J. -M. and {Lasenby}, A. and {Lattanzi}, M. and {Lawrence}, C.~R. and {Le Jeune}, M. and {Lemos}, P. and {Lesgourgues}, J. and {Levrier}, F. and {Lewis}, A. and {Liguori}, M. and {Lilje}, P.~B. and {Lilley}, M. and {Lindholm}, V. and {L{\'o}pez-Caniego}, M. and {Lubin}, P.~M. and {Ma}, Y. -Z. and {Mac{\'\i}as-P{\'e}rez}, J.~F. and {Maggio}, G. and {Maino}, D. and {Mandolesi}, N. and {Mangilli}, A. and {Marcos-Caballero}, A. and {Maris}, M. and {Martin}, P.~G. and {Martinelli}, M. and {Mart{\'\i}nez-Gonz{\'a}lez}, E. and {Matarrese}, S. and {Mauri}, N. and {McEwen}, J.~D. and {Meinhold}, P.~R. and {Melchiorri}, A. and {Mennella}, A. and {Migliaccio}, M. and {Millea}, M. and {Mitra}, S. and {Miville-Desch{\^e}nes}, M. -A. and {Molinari}, D. and {Montier}, L. and {Morgante}, G. and {Moss}, A. and {Natoli}, P. and {N{\o}rgaard-Nielsen}, H.~U. and {Pagano}, L. and {Paoletti}, D. and {Partridge}, B. and {Patanchon}, G. and {Peiris}, H.~V. and {Perrotta}, F. and {Pettorino}, V. and {Piacentini}, F. and {Polastri}, L. and {Polenta}, G. and {Puget}, J. -L. and {Rachen}, J.~P. and {Reinecke}, M. and {Remazeilles}, M. and {Renzi}, A. and {Rocha}, G. and {Rosset}, C. and {Roudier}, G. and {Rubi{\~n}o-Mart{\'\i}n}, J.~A. and {Ruiz-Granados}, B. and {Salvati}, L. and {Sandri}, M. and {Savelainen}, M. and {Scott}, D. and {Shellard}, E.~P.~S. and {Sirignano}, C. and {Sirri}, G. and {Spencer}, L.~D. and {Sunyaev}, R. and {Suur-Uski}, A. -S. and {Tauber}, J.~A. and {Tavagnacco}, D. and {Tenti}, M. and {Toffolatti}, L. and {Tomasi}, M. and {Trombetti}, T. and {Valenziano}, L. and {Valiviita}, J. and {Van Tent}, B. and {Vibert}, L. and {Vielva}, P. and {Villa}, F. and {Vittorio}, N. and {Wandelt}, B.~D. and {Wehus}, I.~K. and {White}, M. and {White}, S.~D.~M. and {Zacchei}, A. and {Zonca}, A.},
        title = "{Planck 2018 results. VI. Cosmological parameters}",
      journal = {Astron. Astrophys.},
         year = 2020,
        month = sep,
       volume = {641},
          eid = {A6},
        pages = {A6},
          doi = {10.1051/0004-6361/201833910},
archivePrefix = {arXiv},
       eprint = {1807.06209},
 primaryClass = {astro-ph.CO},
       adsurl = {https://ui.adsabs.harvard.edu/abs/2020A&A...641A...6P}
}

@ARTICLE{Grav20,
       author = {{Gravity Collaboration} and {Abuter}, R. and {Amorim}, A. and {Baub{\"o}ck}, M. and {Berger}, J.~P. and {Bonnet}, H. and {Brandner}, W. and {Cardoso}, V. and {Cl{\'e}net}, Y. and {de Zeeuw}, P.~T. and {Dexter}, J. and {Eckart}, A. and {Eisenhauer}, F. and {F{\"o}rster Schreiber}, N.~M. and {Garcia}, P. and {Gao}, F. and {Gendron}, E. and {Genzel}, R. and {Gillessen}, S. and {Habibi}, M. and {Haubois}, X. and {Henning}, T. and {Hippler}, S. and {Horrobin}, M. and {Jim{\'e}nez-Rosales}, A. and {Jochum}, L. and {Jocou}, L. and {Kaufer}, A. and {Kervella}, P. and {Lacour}, S. and {Lapeyr{\`e}re}, V. and {Le Bouquin}, J. -B. and {L{\'e}na}, P. and {Nowak}, M. and {Ott}, T. and {Paumard}, T. and {Perraut}, K. and {Perrin}, G. and {Pfuhl}, O. and {Rodr{\'\i}guez-Coira}, G. and {Shangguan}, J. and {Scheithauer}, S. and {Stadler}, J. and {Straub}, O. and {Straubmeier}, C. and {Sturm}, E. and {Tacconi}, L.~J. and {Vincent}, F. and {von Fellenberg}, S. and {Waisberg}, I. and {Widmann}, F. and {Wieprecht}, E. and {Wiezorrek}, E. and {Woillez}, J. and {Yazici}, S. and {Zins}, G.},
        title = "{Detection of the Schwarzschild precession in the orbit of the star S2 near the Galactic centre massive black hole}",
      journal = {Astron. Astrophys.},
         year = 2020,
        month = apr,
       volume = {636},
          eid = {L5},
        pages = {L5},
          doi = {10.1051/0004-6361/202037813},
archivePrefix = {arXiv},
       eprint = {2004.07187},
 primaryClass = {astro-ph.GA},
       adsurl = {https://ui.adsabs.harvard.edu/abs/2020A&A...636L...5G}
}

@ARTICLE{Grav18,
       author = {{Gravity Collaboration} and {Abuter}, R. and {Amorim}, A. and {Anugu}, N. and {Baub{\"o}ck}, M. and {Benisty}, M. and {Berger}, J.~P. and {Blind}, N. and {Bonnet}, H. and {Brandner}, W. and {Buron}, A. and {Collin}, C. and {Chapron}, F. and {Cl{\'e}net}, Y. and {Coud{\'e} Du Foresto}, V. and {de Zeeuw}, P.~T. and {Deen}, C. and {Delplancke-Str{\"o}bele}, F. and {Dembet}, R. and {Dexter}, J. and {Duvert}, G. and {Eckart}, A. and {Eisenhauer}, F. and {Finger}, G. and {F{\"o}rster Schreiber}, N.~M. and {F{\'e}dou}, P. and {Garcia}, P. and {Garcia Lopez}, R. and {Gao}, F. and {Gendron}, E. and {Genzel}, R. and {Gillessen}, S. and {Gordo}, P. and {Habibi}, M. and {Haubois}, X. and {Haug}, M. and {Hau{\ss}mann}, F. and {Henning}, Th. and {Hippler}, S. and {Horrobin}, M. and {Hubert}, Z. and {Hubin}, N. and {Jimenez Rosales}, A. and {Jochum}, L. and {Jocou}, K. and {Kaufer}, A. and {Kellner}, S. and {Kendrew}, S. and {Kervella}, P. and {Kok}, Y. and {Kulas}, M. and {Lacour}, S. and {Lapeyr{\`e}re}, V. and {Lazareff}, B. and {Le Bouquin}, J. -B. and {L{\'e}na}, P. and {Lippa}, M. and {Lenzen}, R. and {M{\'e}rand}, A. and {M{\"u}ler}, E. and {Neumann}, U. and {Ott}, T. and {Palanca}, L. and {Paumard}, T. and {Pasquini}, L. and {Perraut}, K. and {Perrin}, G. and {Pfuhl}, O. and {Plewa}, P.~M. and {Rabien}, S. and {Ram{\'\i}rez}, A. and {Ramos}, J. and {Rau}, C. and {Rodr{\'\i}guez-Coira}, G. and {Rohloff}, R. -R. and {Rousset}, G. and {Sanchez-Bermudez}, J. and {Scheithauer}, S. and {Sch{\"o}ller}, M. and {Schuler}, N. and {Spyromilio}, J. and {Straub}, O. and {Straubmeier}, C. and {Sturm}, E. and {Tacconi}, L.~J. and {Tristram}, K.~R.~W. and {Vincent}, F. and {von Fellenberg}, S. and {Wank}, I. and {Waisberg}, I. and {Widmann}, F. and {Wieprecht}, E. and {Wiest}, M. and {Wiezorrek}, E. and {Woillez}, J. and {Yazici}, S. and {Ziegler}, D. and {Zins}, G.},
        title = "{Detection of the gravitational redshift in the orbit of the star S2 near the Galactic centre massive black hole}",
      journal = {Astron. Astrophys.},
         year = 2018,
        month = jul,
       volume = {615},
          eid = {L15},
        pages = {L15},
          doi = {10.1051/0004-6361/201833718},
archivePrefix = {arXiv},
       eprint = {1807.09409},
 primaryClass = {astro-ph.GA},
       adsurl = {https://ui.adsabs.harvard.edu/abs/2018A&A...615L..15G}
}

@ARTICLE{Do2019,
       author = {{Do}, Tuan and {Hees}, Aurelien and {Ghez}, Andrea and {Martinez}, Gregory D. and {Chu}, Devin S. and {Jia}, Siyao and {Sakai}, Shoko and {Lu}, Jessica R. and {Gautam}, Abhimat K. and {O'Neil}, Kelly Kosmo and {Becklin}, Eric E. and {Morris}, Mark R. and {Matthews}, Keith and {Nishiyama}, Shogo and {Campbell}, Randy and {Chappell}, Samantha and {Chen}, Zhuo and {Ciurlo}, Anna and {Dehghanfar}, Arezu and {Gallego-Cano}, Eulalia and {Kerzendorf}, Wolfgang E. and {Lyke}, James E. and {Naoz}, Smadar and {Saida}, Hiromi and {Sch{\"o}del}, Rainer and {Takahashi}, Masaaki and {Takamori}, Yohsuke and {Witzel}, Gunther and {Wizinowich}, Peter},
        title = "{Relativistic redshift of the star S0-2 orbiting the Galactic Center supermassive black hole}",
      journal = {Science},
         year = 2019,
        month = aug,
       volume = {365},
       number = {6454},
        pages = {664-668},
          doi = {10.1126/science.aav8137},
archivePrefix = {arXiv},
       eprint = {1907.10731},
 primaryClass = {astro-ph.GA},
       adsurl = {https://ui.adsabs.harvard.edu/abs/2019Sci...365..664D}
}



\end{document}